\documentclass[12pt]{article}
\usepackage[backref]{hyperref}
\usepackage{lineno} 
\usepackage{natbib} 
\usepackage{amsmath}
\usepackage{amssymb}
\usepackage{amsthm} 
\usepackage{verbatim}
\usepackage[T1]{fontenc}
\usepackage{longtable} 
\usepackage{graphicx}
\usepackage{adjustbox}
\usepackage{multirow}
\usepackage{hyperref}
\usepackage{caption}
\usepackage{subcaption}
\usepackage{float}
\usepackage{placeins}
\usepackage{bbm}
\usepackage{bm}
\usepackage[american]{babel}
\usepackage{enumitem}
\setlist{leftmargin=5mm}

\usepackage{tikz} 
\usetikzlibrary{shapes,arrows}
\usetikzlibrary{automata,positioning}
\tikzstyle{block} = [draw, fill=white, rectangle, 
    minimum height=3em, minimum width=7em]
\tikzstyle{input} = [coordinate]
\tikzstyle{output} = [coordinate]

\usepackage{xcolor}
\definecolor{darkgreen}{rgb}{0.0, 0.5, 0.0}
\usepackage[linesnumbered,ruled,vlined]{algorithm2e}

\SetCommentSty{mycommfont}

\newcommand{\blind}{1}

\begin{document}

\def\spacingset#1{\renewcommand{\baselinestretch}%
{#1}\small\normalsize} \spacingset{1}


\if1\blind
{
  \title{\bf Nonparametric robust monitoring of time series panel data}
  \author{Sophie Mathieu\thanks{
    The authors gratefully acknowledge \textit{funding from the Belgian Federal Science Policy Office (BELSPO) through the BRAIN VAL-U-SUN project (BR/165/A3/VAL-U-SUN) and the computational resources provided by the Consortium des {\'E}quipements de Calcul Intensif (CECI), funded by the Fonds de la Recherche Scientifique de Belgique (F.R.S.-FNRS) under Grant No. 2.5020.11 and by the Walloon Region.}}\hspace{.2cm}\\
    LIDAM, UCLouvain\\
    and \\
    Rainer von Sachs \\
    LIDAM, UCLouvain \\
		    and \\
    V{\'e}ronique Delouille \\
    Solar Physics and Space Weather department, Royal Observatory of Belgium\\
		    and \\
    Laure Lef{\`e}vre \\
     Solar Physics and Space Weather department, Royal Observatory of Belgium \\
		    and \\
    Christian Ritter \\
    LIDAM, UCLouvain}
  \maketitle
} \fi

\if0\blind
{
  \bigskip
  \bigskip
  \bigskip
  \begin{center}
    {\LARGE\bf Nonparametric robust monitoring of time series panel data}
\end{center}
  \medskip
} \fi

\bigskip

\begin{abstract}
In many applications, a control procedure is required to detect potential deviations in a panel of serially correlated processes. It is common that the processes are corrupted by noise and that no prior information about the in-control data are available for that purpose. 
This paper suggests a general nonparametric monitoring scheme for supervising such a panel with time-varying mean and variance. The method is based on a control chart designed by block bootstrap, which does not require parametric assumptions on the distribution of the data.  
The procedure is tailored to cope with strong noise, potentially missing values and absence of in-control series, which is tackled by an intelligent exploitation of the information in the panel. 
Our methodology is completed by support vector machine procedures to estimate magnitude and form of the encountered deviations (such as stepwise shifts or functional drifts).
This scheme, though generic in nature, is able to treat an important applied data problem: the control of deviations in a subset of sunspot number observations which are part of the International Sunspot Number, a world reference for long-term solar activity. 
\end{abstract}

\noindent%
{\it Keywords:}  Statistical process control; Support vector machine; Sunspot numbers; Correlation; Noise; Quality control
\vfill

\newpage
\spacingset{1.45} 
\section{Introduction}
\label{sec:intro}

It is a common problem in many applications to continuously monitor the mean of a process in order to detect as soon as possible any deviation from its well-performing level. It is recurrent in industry, where most of the production lines need to be under permanent control to reduce the number of production defects and ensure a standard product quality. Similar problems also appear in various fields such as manufacturing, aeronautics or medicine. \\
 This paper proposes a general nonparametric method to monitor the mean level of a panel of processes corrupted by strong noise. No a-priori knowledge about the data, such as information about the non-deviating or in-control (IC) and deviating or out-of-control (OC)  processes, the distributions of the observations or the nature of their correlation, is required. 
Therefore, the method can be used with complex data that are heavily affected by several kinds of deviations (ranging from sudden jumps to oscillating shifts). 
 
\subsection{Motivations}
 This paper is driven by a particular application in astrophysics, related to one of the longest-running scientific experiment~\citep{Owens2013}: the observation of the sunspots. 
The sunspots are dark regions on the Sun associated to a high local magnetic field. They have been observed and counted since the seventeenth century in different observatories, also called ``stations'', across the world and serve as an international reference for the long-term solar activity~\citep{Ermolli2013}. 
 To keep the continuity on how the data have been collected, the observations are still performed today with similar methods as those used at the birth of the series, a few centuries ago. These observation methods, combined with the intrinsic differences between the stations (weather conditions, instruments etc), lead to a panel of series which are corrupted by strong noise. Hence, the data are subject to many deviations along time~\citep{Clette2016, all_corrections}. They are also serially correlated and non-normally distributed with an important amount of missing values, due to weather conditions preventing the observation of the Sun. 
It is therefore crucial to develop an effective monitoring for supervising the quality of this panel of complex series and for controlling their long-term stability. 

\subsection{Existing control schemes and limitations}

To this end, \cite{Qiu2014} develop a dynamic screening system based on extensions of the classical CUSUM~\citep{Page1961} chart. 
Those extensions include a nonparametric design of the chart based on a block bootstrap procedure to construct a scheme robust to non-normally distributed and serially correlated data. In their method, the regular IC patterns are also estimated using cross-sectional information over a panel of IC processes. 
These information allow the chart to detect shifts in the mean level of each process, where the process is allowed to have a mean changing over time. The procedure also allows the monitoring of processes which do not have an IC period. 
However, the method proposed in~\cite{Qiu2014} cannot be directly applied on data which deviate much over time: 
\begin{enumerate}
\setlength{\itemsep}{0.1cm}
\item The scheme cannot be used without knowing from the beginning which processes are in-control. Those information is not available in many applications such as the sunspot number data, where \emph{all} series are expected to contain several kinds of deviations 
 along their observation period.
Therefore, we propose in the following a procedure which better exploits the information in the panel. 
Contrarily to previous work where the set of series is only used to construct a reliable distribution of the data under the IC regime, we also benefit from the panel to establish an IC reference that is robust against the influence of potentially deviating individual processes. 
This reference allows us to monitor a panel of processes without presumptive information about the data. 

\item The method operates with a control chart which sends an alert when a deviation is detected, yet without providing any information about the nature of the shifts. Such information is however valuable since it allows different responses as a function of the magnitude or the form of the shift (ranging from simple warnings to the immediate interruption of the process). 
Although several methods have been developed in the literature to predict automatically the size of shifts after an alert, see for instance~\cite{Cheng2011} and its bibliography, they are not adapted to cases where the observations are (simultaneously) non-normally distributed, serially correlated and contaminated by strong noise.  In the following, we will thus propose a method 
that predicts efficiently the characteristics (i.e. the magnitude and the form) of the shifts in those more general conditions. 
\end{enumerate}

\subsection{Contributions}
In this paper, we build upon the existing statistical process control (SPC) literature and extend it to propose a generic methodology for monitoring a panel of observations corrupted by strong noise. 
Although motivated by a particularly complex dataset, the methodology is general and can be applied to other data. 
The method integrates different techniques, ranging from classical control charts to more recent machine learning procedures, 
 to present a complete monitoring scheme: this starts with the question of how to select the IC observations based on a robust criterion involving the whole time series panel, extends to developing an efficient control chart and finally provides estimations of the characteristics of the shift for each alert. All existing methods have not only been assembled but also modified to work efficiently for a panel of autocorrelated and non-normally distributed observations, prone to a potentially high number of both missing values and outliers.
Moreover, contrarily to what is classically treated in the literature, the panel is generically constructed in a multiplicative, rather than additive, framework.  \\
This article is structured as follows. In Section \ref{sec:method}, we present our model and our complete methodology. 
In Section \ref{sec:appli}, we apply the proposed method to the particular sunspot number data. It turns out that our method is capable to automatically detect significant deviations in some prominent observing stations that had been identified in the past by human comparisons as well as many others undetected.
 In a final section \ref{sec:conclusion}, we give some concluding remarks. 
An appendix section is also devoted to present some omitted details of the different procedures used to build up our methodology. 

\section{Methodology}
\label{sec:method}

\begin{figure}[H]
\begin{tikzpicture}[auto, node distance=3.5cm, >=latex']

    \node (input) [block, text width=2.5cm, align=center, dashed, label=Ia] {Selection of IC processes};
    \node (fft) [block, right of=input, text width=2.5cm, align=center, label=Ib] {Estimation of IC patterns};
    \node (ths) [block, right of=fft, text width=2.6cm, align=center, label=IIa] {Design of the chart by BB}; 
    \node (ifft) [block, right of=ths, text width=2.5cm, align=center, label=IIa] {Monitoring of standardized data};
    \node (iasc) [block, right of=ifft, text width=2.6cm, align=center, dashed, label=IIb] {Estimation of shift sizes/forms by SVMs};
      
     \draw [->] (input) -- node[midway, above] {} (fft);
     \draw [->] (fft) -- node[name=u3] {} (ths);
     \draw [->] (ths) -- node[name=u4] {} (ifft);
     \draw [->] (ifft) -- node[name=u5] {} (iasc);
		
	 \node[below of=input, node distance=1.7cm, font=\itshape] {cal.};
	\node[below of=fft, node distance=1.7cm, font=\itshape] {cal.};
	\node[below of=ths, node distance=1.7cm, font=\itshape] {cal.};
       
\end{tikzpicture}
\caption{\footnotesize{ General framework of the methodology. It is in particular in the dashed blocks that we contribute with completely new ingredients of the method. ``cal.'' denotes the offline calibration stages of the procedure whereas ``I'' and ``II'' refer respectively to the phase I and phase II SPC. Note that the calibration also contains the training (and validation) steps of the SVM procedures. These steps are explained in the phase IIb for clarity, since they are not included in a classical monitoring scheme.  
}}
 \label{Fig:method}
\end{figure}
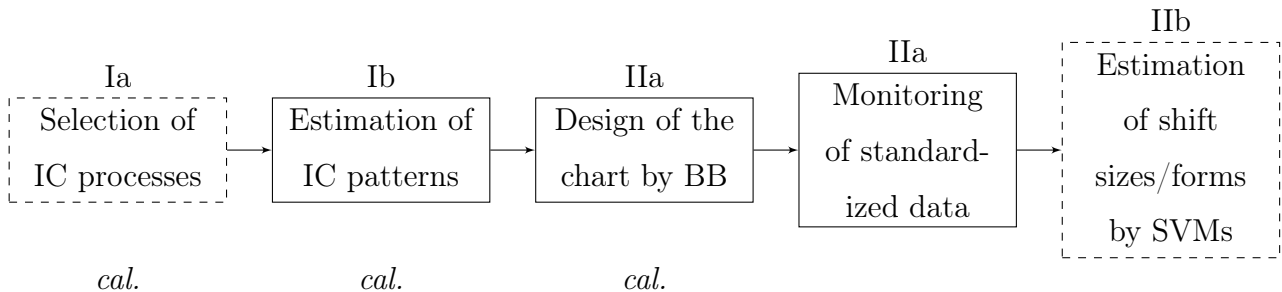

In this section, our complete methodology is explained after presenting the model.
In Phase Ia of SPC, we select a subset of IC processes from the panel. Then, we estimate in Phase Ib the IC longitudinal patterns of the series. In Phase IIa, data are standardized by the estimated IC patterns and monitored by a control chart. 
We first explain how the chart would be designed for independent and identically distributed (i.i.d.) non-normal data, then we introduce the block bootstrap (BB) approach to adapt the method to serially correlated data.
Finally,  we estimate the size and the form of the shifts in Phase IIb, using support vector machine (SVM) procedures on top of the control chart. \\
In practice, after an offline calibration on past (available) observations, the procedure is applied in real-time for online monitoring, as described in Figure \ref{Fig:method}. 
If the control scheme is used for a long time, the panel of processes as well as the individual processes may evolve. 
As a result, the calibration may be periodically rerun, when we suspect that the panel of processes has significantly changed. 

\subsection{Model}
\label{sec:model}

Assume that we are interested in monitoring a panel of processes $X(i,t)$, where $i$, $i \in 1,..., N$, denotes the index of the process and $t$, $t \in 1,...,T$, represents the observation time.
The processes are supposed to be equally-spaced and may contain missing values. 
In the simplest approach, the different processes may be written in an additive model as:
\begin{equation}
\label{E:model-add}
\begin{split}
X(i,t) = \left\{ \begin{array}{ll} c(t) + \eta(i,t) + h(i,t) & \text{for} \  t \in [0, \tau] \\ 
c(t) + \eta(i,t) + f(\delta; i,t) + h(i,t) & \text{for} \  t \in [\tau, T], \\ \end{array} \right. 
\end{split}
\end{equation}
where $c(t)$ represents a signal common to all processes and $\eta(i,t)$ is a process-dependent quantity that experiences a deviation $f(\delta; i,t)$ of magnitude $\delta$ at time $t=\tau$.   
Moreover, if the processes have an intrinsic level related to an external factor that is not relevant to control, the model may also contain a variable $h(i,t)$.  We assume that this level $h(i,t)$ is a slowly-changing function, which does not vary too much with respect to $\eta(i,t)$. \\
To detect the deviation $f(\delta; i,t)$, the influence of the common signal and the intrinsic level should be removed from the observations. 
To that end, we suggest to estimate the common signal $c(t)$ by the median $\hat c(t)$ of the processes across the panel, which is a robust estimator against individual deviations, and subtract $\hat c(t)$ from the observations: $\hat{ \widetilde{\eta}}(i,t) =X(i,t)-\hat c(t)$. \\

W.l.o.g., we focus in the following on the more complex situation where the processes follow a multiplicative framework:
\begin{equation}
\label{E:model}
\begin{split}
X(i,t) = \left\{ \begin{array}{ll} c(t) (\eta(i,t) + h(i,t)) & \text{for} \  t \in [0, \tau] \\ 
c(t) (\eta(i,t) + f(\delta; i,t) + h(i,t)) & \text{for} \  t \in [\tau, T]. \\ \end{array} \right.  
\end{split}
\end{equation}
The common signal may then be removed from the processes by \emph{dividing} the observations by $\hat c(t)$:
\begin{equation}
\label{E:error}
\hat{ \widetilde{\eta}}(i,t) =\frac{X(i,t)}{\hat c(t)}=\frac{X(i,t)}{\underset{1 \leq i\leq N}{\text{med}} X(i,t)}.
\end{equation}
Note that $\widetilde{\eta}(i,t) =\eta(i,t) + h(i,t)$ is a ratio, it thus depends on the level of the signal. 
However, this effect will not affect the monitoring since the proposed methodology is robust to changes in the mean and the variance of the processes.  

After eliminating the common signal, the individual levels of the processes, if any, may also be removed. 
Since $h(i,t)$ is a slowly-changing function with respect to $\eta(i,t)$, we estimate the levels by applying a smoothing process in time by a sliding window (i.e. an MA-filter) on $\hat{ \widetilde{\eta}}(i,t)$ and subtract them from the processes:
\begin{equation}
\label{E:level}
\begin{split}
 \hat{{\eta}}(i,t)=\hat{ \widetilde{\eta}}(i,t) - \hat{ \widetilde{\eta}}(i,t)^\star,
\end{split}
\end{equation}
where the $\star$ denotes the MA-filter with a sufficiently long window. This window should be chosen depending on the problem at hand. To avoid introducing a bias, it should be the same for all series.

\subsection{Phase I SPC: Estimation of the regular IC longitudinal pattern}

\subsubsection{Phase Ia: Selection of the IC processes}
\label{sec:Ia}

If some a-priori information about the processes are established (from previous studies or physical expertise), this knowledge may be used to select a set, called ``pool'', of IC processes from the panel. 
Otherwise,  we propose to select the pool using a stability criterion such as the mean squared error (MSE) with respect to a robust reference of the network: the median of the panel. 
 To have a small MSE, a series should therefore be aligned with the median most of the time and should have a small variance. 
Different clustering algorithms may then be employed, using the values of the MSEs to split the processes into two groups: IC or OC. Such algorithms, generating exactly two groups, include the popular $k$-means~\citep{Lloyd1957, MacQueen1967} or a similar method based on variable Gaussian clusters: the expectation-maximization clustering using Gaussian mixture model~\citep{Dempster1977}, which is expected to perform better than $k$-means when the clusters are not symmetric.
As a result, a pool denoted $P_1$ and containing $N_{IC}$ (determined by the clustering method) processes is automatically generated. 
$P_1$ is stable enough to calibrate the control limits of the chart. However, $P_1$ contains too many intermediately stable processes to be used for accurate estimation of the IC patterns of the data. Therefore, before calibrating the chart, a smaller set of $N_{\widetilde{IC}}$ very stable processes, denoted by $P_2$, is constructed by applying the same clustering method as before on $P_1$ (instead of on the whole panel).
Both $P_2$ and even more $P_1$ contain deviations, called \emph{disparities} to be distinguished from the deviations that are supposed to be actually detected by the method. These \emph{disparities} are expected to be typically smaller and less frequent than the deviations occurring in the processes outside of $P_1$.

In some particular cases such as our application, $P_1$ may in addition suffer from deviations that are of similar magnitude as those of the OC processes. To cope with this and preserve the detection power of our scheme, we apply in these cases a Shewhart chart~\citep{Shewhart1931} with \emph{adaptive} confidence intervals: we remove the IC observations that do not fall into a multiple of the interquartile range (IQR) around the median ($\underset{1 \leq i\leq N}{\text{med}} \hat{ \eta}(i,t)$). 
Note that this adaptive Shewhart would not be a substitute for our control scheme: it only removes the largest deviations at each time without taking into account the history of the process. Therefore, contrarily to our method, it cannot detect the small and persistent shifts.
Thereafter, $P_2$ will be used to accurately estimate the IC patterns of the data, whereas the chart will be calibrated on $P_1$. 
In that way, the chart will properly designed. It will also be robust to the disparities of $P_1$, which otherwise, for the applications at hand, would turn to be over-sensitive. 


\subsubsection{Phase Ib: Estimation of the mean and the variance of the IC processes}
\label{sec:Ib}

The empirical mean and variance, denoted respectively by $\hat \mu_0(t)$ and $\hat \sigma_0^2(t)$ are then calculated on the IC processes:
\begin{equation}
\label{E:musigma}
\begin{split}
 \hat \mu_0(t)=\frac{1}{\Delta} \sum_{t'=t-\Delta /2}^{t+\Delta /2} \frac{1}{N_{\widetilde{IC}}} \sum_{i_{\widetilde{ic}}=1}^{N_{\widetilde{IC}}} \hat \eta (i_{\widetilde{ic}}, t') \\
 \hat \sigma_0^2(t)=\frac{1}{\Delta} \sum_{t'=t-\Delta /2}^{t+\Delta /2} \frac{1}{N_{\widetilde{IC}}} \sum_{i_{\widetilde{ic}}=1}^{N_{\widetilde{IC}}} \left(\hat \eta (i_{\widetilde{ic}},t') - \hat \mu_0(t) \right)^2, 
\end{split}
\end{equation}
where $i_{\widetilde{ic}}$ denotes the index of a station included in $P_2$. 
Note that we only give boxcar (rectangular kernel) smoothing in (\ref{E:musigma}), although any generic window smoothing with a smoother kernel works just as well. 
When there are missing values (such as in our application), the IC patterns may be computed using nearest neighbours (K-NN) estimators. 
Hence, the temporal information may compensate the higher rates of missing observations that some processes experience.  

In the calibration phase, a symmetric window around $t$ may be used to compute $\hat \mu_0(t)$ and $\hat \sigma_0(t)$ for each new observations, since all data are already available. 
In real-time however, the temporal window needs to be left-sided to include only the past and current observations. 

\subsection{Phase II SPC: Monitoring of the longitudinal pattern of the observations}
The observations of all processes are then standardized by the estimated IC patterns: 
\begin{equation}
\label{E:residuals}
\hat \epsilon_{\hat \eta}(i,t) =\frac{(\hat \eta(i,t) -\hat \mu_0(t))} {\hat \sigma_0(t)} 
\end{equation}

From now on, we omit the index $i$ since each process will be monitored separately. 
A classical control chart such as the cumulative sum (CUSUM)~\citep{Page1961} or the exponentially weighted moving average (EWMA)~\citep{Roberts1959} may then be applied on the residuals. 
W.l.o.g., the two-sided CUSUM chart applied on the residuals writes as:
\begin{equation}
\label{E:CUSUM}
\begin{split}
C_j^+= max(0,C_{j-1}^+ + \hat \epsilon_{\hat \eta}(t)-k) \\
C_j^-= min(0,C_{j-1}^- + \hat \epsilon_{\hat \eta}(t)+k),
\end{split}
\end{equation}
where  $j \geq 1$, $C_0^+ =C_0^-=0$ and $k>0$ is the allowance parameter.\\
This chart gives an alert if  $C_j^+> h^+$ or $C_j^-< h^-$, where $h^-$ and $h^+$ are the control limits of the chart. If the distribution of the residuals is symmetric, $h^-=-h^+$. 

\subsubsection{Phase IIa: Design of the chart for i.i.d. non-normal data}

We first assume that the data are i.i.d., a more general case where the data are serially correlated is treated in the following subsection. It often happens that we aim to monitor data with unknown IC distribution. 
In this case, the CUSUM chart may be calibrated on $P_1$ as follows. 
The allowance parameter, $k$, is first specified in advance. It may be fixed to $k=\delta/2$, where $\delta$ represents the size of the shift that we aim to detect~\citep{Moustakides1986}.
The control limits of the chart are then adjusted by a searching algorithm until a pre-specified rate of false positive (evaluated using the IC average run length) is reached with a desired accuracy. 
This algorithm, explained in details in Appendix \ref{sec:AA}, is based on the bootstrap procedure that randomly samples data with repetitions from the IC residuals. 
Cases where the relation $k=\delta/2$ is suboptimal or when an adequate value for $\delta$ is unknown are also discussed in the appendix.

\subsubsection{Phase IIa: Design of the chart for autocorrelated data}

In general, the residuals may be serially correlated. 
If the autocorrelation of the data is sufficiently simple, it may be represented by an auto-regressive and moving average (ARMA) model~\citep{time_series}. 
 However, in the most general case, an ARMA model is inadequate to model complex autocorrelation structures. 
To that end, the bootstrap procedure may be modified to take into account the actual autocorrelation of the data. 
Instead of sampling a sequence of unique observations with repetitions from the IC residuals, these data are sampled per block, which are then concatenated to form a complete series of the desired length. This method, called the block bootstrap (BB), preserves the serial correlation of the data inside the blocks. 
As theoretically demonstrated in~\cite{Lahiri1999}, BB methods using non-overlapping blocks and random block lengths are more variable than those based on overlapping blocks and constant lengths. Therefore, we select the popular well-studied moving BB (MBB)~\citep{Kunsch1989, Liu1992}, to obtain the best performances. 
As an illustration of the BB method, we compare a classical approach, where the autocorrelation of the data is removed by fitting an ARMA model, to the BB to calibrate the chart for non-trivially serially correlated data in Appendix \ref{sec:AB}. 
The matched BB~\citep{Carlstein1998} may also be implemented for data experiencing long-term autocorrelations. \\
We also mention that the data may be decorrelated by a sequential scheme based on recursive covariance matrix decomposition~\citep{QiuLi2016, QiuYou2018}. According to the latter papers, this method shows the same results as the BB as long as the sampling rate is the same in the IC and OC data, which is by assumption true in this paper. 
However, the method can be computationally intensive since the data are decorrelated until the last re-start of the CUSUM chart, which can extend back in time for long series prone to a high number of deviations. \\ 
In case of missing values, we can still apply the bootstrap procedures previously described, to calibrate the chart. If the missing values are independent of the observations, the chart may be restarted at each missing values ($C_j^+ =C_j^-=0$). When many missing observations are expected, the chart statistics may also be propagated through the shortest gaps (and only reset for large gaps). This is motivated by the fact that the short and large gaps (to be defined according to the application) usually originate from different phenomena.  
Another option is to use a more complex chart such as an EWMA in the framework of likelihood ratio test (denoted by EWMA-GLR in the following)~\citep{Qiu2017}, that can automatically handle missing values.

\subsection{Phase IIb: Estimation of the sizes and forms of the shifts using SVMs}
Traditional control charts give an alert when a deviation is detected in the data, yet without providing any information about the characteristics (forms and sizes) of the shift. This knowledge is however valuable to assign possible causes to the shifts or to adapt the type of alert that is sent back. 
To that end, ~\cite{Cheng2011} appended a support vector regression (SVR) to the CUSUM chart to predict the size of the deviations for i.i.d. normally distributed data.
In this paper, we extend~\cite{Cheng2011} to render the method effective to detect the characteristics of the deviations in serially correlated and non-normal data with missing values. This includes the addition of an SVM classifier (SVC) of the different forms of the shifts from a finite predefined set of functions. \\
We denote by $\delta_{min}$ the size of the smallest shift that we aim to detect. 
To identify the characteristics of the shifts, an SVR and an SVC are added on top of the control chart.
When the chart detects a deviation, the last observations are fed into the SVR and SVC, previously trained on a wide range of simulated shifts. 
Then, the SVR predicts the magnitude of the deviations equal to or greater than $\delta_{min}$ with: 
\begin{equation}
\label{E:SVR}
\hat \delta=f(V_{\tau})=f(\hat \epsilon_{\hat \eta}(\tau-m+1),\hat \epsilon_{\hat \eta}(\tau-m+2),...,\hat \epsilon_{\hat \eta}(\tau)),
\end{equation}
where $\tau$ denotes the time of the alert and $V_{\tau}$ represents the input vector, i.e. a sequence containing the $m$ last observations of the process.  
The SVC also estimates the form of the shifts on the same input vector.
We refer to Appendix \ref{sec:AC} for more insights into the SVM procedures. 
The length $m$ of $V_{\tau}$ is an important parameter of the method. It should therefore be carefully selected using one of the methods described in Appendix \ref{sec:AC1}. 
In cases of missing observations, it is likely that $V_{\tau}$ will contain gaps. In this case, we discard the input vectors which do not contain at least 20\% of valid observations. Missing observations occurring at the beginning of $V_{\tau}$ are simply replaced by the first valid observation encountered, while the ``intermediate'' gaps are filled by a linear interpolation. 
Note that, since the input vector is constructed from the residuals and not from the CUSUM statistics, this method can be applied in general with any control chart. 

\subsubsection{Creation of the training and testing sets}
Since the SVR and SVC are supervised machine-learning procedures, they require appropriate training and testing sets to correctly operate.
Contrarily to what is done in~\cite{Cheng2011}, where the authors simulated simple jumps of size uniformly distributed over a finite set, we modify the design of the instances to accommodate the complex features of the data as follows. 
{\emph{Shift sizes}:} We generate training instances by BB for different shifts, whose magnitudes (the labels of the instances for SVR) are randomly sampled from two half-normal distributions~\citep{distributions} supported by $[-\infty, ...,-\delta_{min}]$ and $[\delta_{min},...,\infty]$ respectively. 
This sampling generates shifts in a continuous range, to better mimic the actual deviations. 
{\emph{Shift forms}:} Different forms of the shifts (the labels of the instances for SVC) are also implemented ranging from simple jumps, to highly-oscillating functions.  
{\emph{Time of the shift}:} In an actual monitoring, the shifts may happen not immediately but after an initial IC period. Thus, we also start the monitoring after a random delay to train the methods at identifying the shifts appearing at any time inside the window length. 

\section{Application}
\label{sec:appli}
In this section, we apply the previously described methodology to the dataset which inspires this research: the number of sunspots. 
The data and their model are first presented in Section \ref{sec:appli_data} and \ref{sec:appli_model}, respectively. Phase I SPC is introduced in Section \ref{sec:appli_I}, where the regular IC pattern of the data are estimated. Section \ref{sec:appli_II} is dedicated to the phase II SPC. 
After comparing several charts, the selected control scheme is adjusted on the standardized data. An SVR together with an SVC are also trained to predict the characteristics of the deviations. 
The method is finally tested on actual OC stations in Section \ref{sec:results}. The procedure automatically detects major deviations identified recently by hand for particular stations. It also unravels small and persistent shifts that are more difficult to identify manually.

\subsection{Data}
\label{sec:appli_data}

Sunspots are seen as dark areas on white light images and correspond to regions with a high local magnetic field. The number of spots ($N_s$)  are counted daily in tens of stations across the world together with the number of sunspot groups ($N_g$). Both numbers are the building blocks of the world reference for long-term solar activity~\citep{Ermolli2013}: the international sunspot number ($S_n$\footnote{The $S_n$ is distributed through the World Data Center Sunspots Index and Long-term Solar Observations (WDC-SILSO) \url{http://www.sidc.be/silso/}}). 
To preserve the uniformity of the series, the observations are still performed today with similar methods (e.g. instruments, counting procedure) as those used at the birth of the series, centuries ago. 
It is therefore essential to develop a powerful monitoring procedure for supervising the quality of these observations and for controlling their long-term stability. Such a scheme should be sufficiently robust to accommodate the complex features of the data including:
(a) missing observations, (b) non-normality, (c) the complex autocorrelation of the series (since the sunspots stay between few hours to few months on the Sun) and (d) the absence of clear in-control periods due to the different kinds of deviations that the stations experience over time.\\
In the following, the monitoring procedure is applied to $N_s$, the number of sunspots. 
This panel of series contains observations from a subset of 21 observatories whose main characteristics are detailed in Table 1 of \cite{Mathieu2019}. 
As in the latter paper, the period under study extends from January 1, 1947, till December 31, 2013. 
To preserve notational consistency, we denote by $t$,  $t \in 1,...,T$, the time of the observation and represent the index of the stations by $i$, $i \in 1,..., N=21$. 
Note that, the number of observations exceeds the number of time series by a factor one hundred, even with the complete database. 

\subsection{Model}
\label{sec:appli_model}

In the spirit of (\ref{E:model}), we establish a \emph{multiplicative} model where 
the observations, $Y_i(t)$, may be decomposed into a common solar signal $s(t)$ corrupted by three types of station-dependent errors: 
\begin{equation} 
\begin{split}
Y_i(t) = \left\{ \begin{array}{ll} (\epsilon_1(i,t)+\epsilon_2(i,t))s(t) & \text{if} \  s(t)>0 \\ 
\epsilon_3(i,t) & \text{if} \  s(t)=0. \\ \end{array} \right.
\end{split}
 \label{E:modelsun}
 \end{equation}
$\epsilon_1$ denotes the short-term error representing counting errors and variable seeing conditions; 
$\epsilon_2$ is a long-term error accounting for systematic bias in the counting. 
The error $\epsilon_3$ is occurring only at solar minima (when $s(t)$ is equal to zero); it captures effects like short-duration sunspots and the non-simultaneity of the observations across the network of stations. \\
We aim at monitoring the long-term error of the stations, $\epsilon_2$,  which represents the potential bias of the observatories.  To this end, we can apply a method similar to those described in Section \ref{sec:model}. However, the procedure is in practice more complex due to the fact that: (a) more stations tend to rather overestimate the number of spots than to underestimate it and (b) the short-term error is entangled with the long-term error and the individual levels of the stations in (\ref{E:modelsun}). To cope with (a), the median is computed on observations $Y$ that are slightly modified as described in~\cite[Section~6.3]{Mathieu2019}. To cope with (b), a smoothing process by a MA-filter with a window length of 27 days is first applied on the observations. This value of 27 days is equal to one solar rotation, a physical scale of the data that is sufficiently high to overcome the effects of the short-term regime. Then, the levels of the stations, which correspond to differences of weather conditions and instruments, are separated from the long-term error using (\ref{E:level}), by applying a MA-filter with a window length equal to 240 days. This value seems appropriate since the location of the observatories or their telescope are unlikely to change much over time (we refer to~\cite[Appendix~A]{Mathieu2019} for a formal derivation of the window length). 
\begin{figure}[hbt]
	\centering
		\includegraphics[scale=0.5]{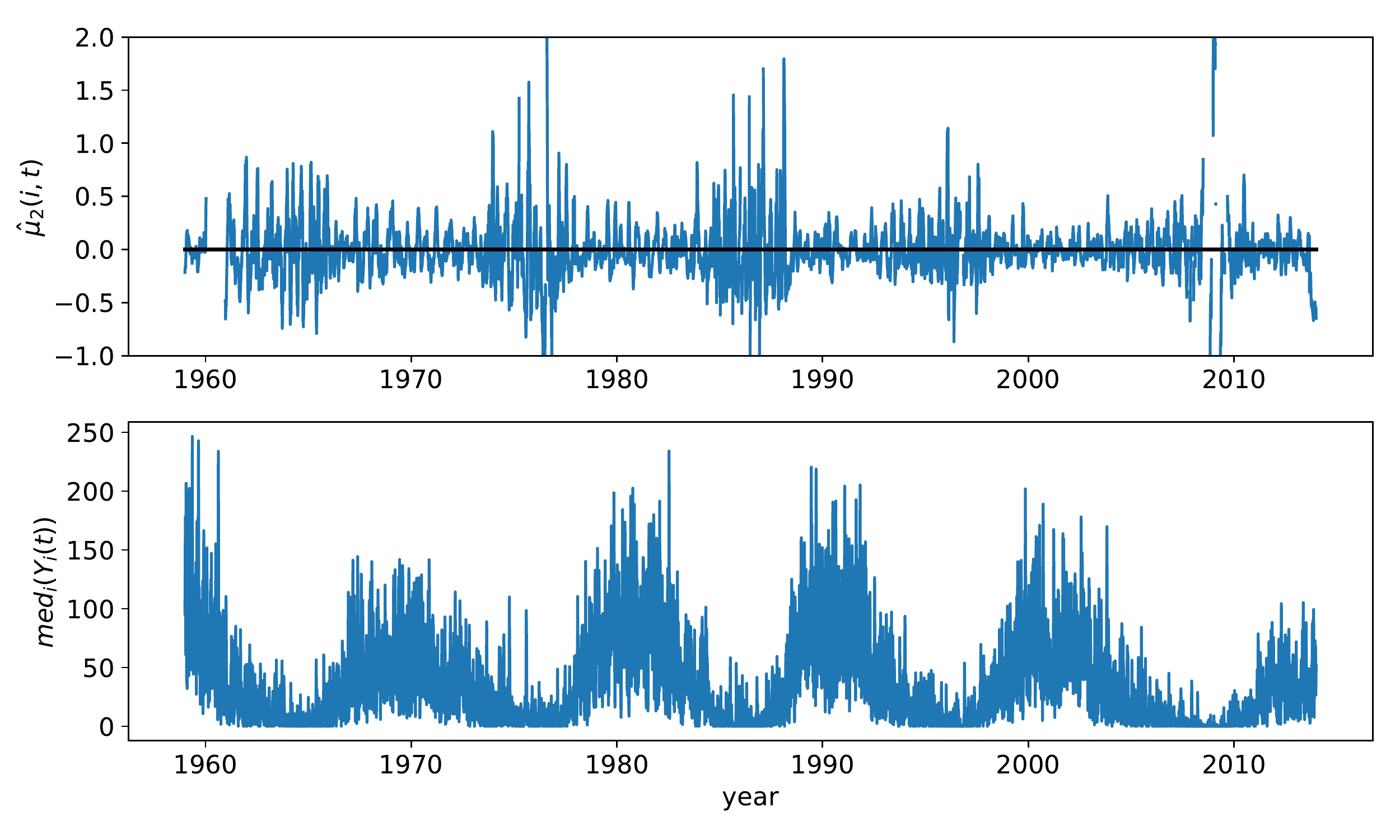}
		\caption{\footnotesize{Long-term error, $\hat \mu_2(i,t)$, in the station Locarno (Switzerland) over its observing period. The median of the observations is also represented below as a robust estimation of the actual number of spots. This figure clearly shows the eleven-years solar cycle~\citep{Hathaway2010} that is intrinsic to the signal. }}
	\label{fig:cycle}
\end{figure}
We denote by $\hat \mu_2(i,t)$ the estimator of the mean of the long-term error (without individual levels), which corresponds to $ \hat{{\eta}}(i,t)$ introduced in the previous section. As an illustration, the $\hat \mu_2$s of a particular station are represented in Figure~\ref{fig:cycle}. 

\subsection{Phase I SPC}
\label{sec:appli_I}

In Phase Ia, an IC pool is first selected using a robust version of the MSE with respect to the median of the network:
\begin{equation}
\label{E:msesun}
MSE[\hat \mu_2](i) = \underset{1 \leq t\leq T}{\text{med}}[\hat \mu_2(i,t)]^2 + \underset{1 \leq t\leq T}{\text{iqr}} \hat \mu_2(i,t),
\end{equation}
since the data are centred around zero by (\ref{E:error}) and (\ref{E:level}).
We denote by $\underset{1 \leq t\leq T}{\text{iqr}} \hat \mu_2(i,t)$ and $\underset{1 \leq t\leq T}{\text{med}} \hat \mu_2(i,t)$ respectively the IQR and the median of the $\hat \mu_2(i,t)$ in each station. 
As explained in Section \ref{sec:Ia}, we select a moderately robust pool, $P_1$, containing 12 stations and a very robust pool, $P_2$, of 6 stations with the $k$-means clustering. 
We also remove from $P_1$ the IC observations which do not fall into one IQR (containing 50\% of the data) around the median over stations $\underset{1 \leq i\leq N}{\text{med}} \hat \mu_2(i,t)$. 
In phase Ib, the IC regular patterns are computed using K-NN estimators in (\ref{E:musigma}) on $P_2$. 
The number $K$ of nearest neighbours is selected at 200, as a trade-off between the bias and the variance of the estimators.  

\subsection{Phase II SPC}
\label{sec:appli_II}

As in (\ref{E:residuals}), the observations of the stations are then standardized by the IC patterns: 
\begin{equation}
\label{E:residuals_sun}
\hat \epsilon_{\hat \mu_2(i,t)} =\frac{\hat \mu_2(i,t)-\hat \mu_0(t)} {\hat \sigma_0(t)} 
\end{equation}
We select the widely-used CUSUM chart for monitoring the data. Although the EWMA and EWMA-GLR show similar results in most of the cases (except for specific shifts where they sometimes outperform the CUSUM), the CUSUM chart has an optimal relation between the allowance parameter and the size of the shift, which facilitates the design of the scheme. This chart also appears to be more 'user-friendly' for a typical observer without much statistical background than more advanced charts such as the EWMA-GLR. 
As the aim of this article is not to compare several charts, we do not show the comparison results here. Other charts may of course be selected for different applications.\\
Since the data are smoothed by a MA-filter of 27 days, the shortest gaps are mainly suppressed from the observations. 
Hence, the CUSUM chart may be reset at each missing values and adjusted on the residuals using the BB procedure with a block length also equal to 27, a physical scale of the data. This value is also selected as a trade-off between an optimal representation of the autocorrelation and the mean and the variance of the bootstrapped data. 
Moreover, we train an SVR with a radial basis function kernel, $\lambda=10$, $\epsilon=0.001$ (see appendix \ref{sec:AC} for more explanations) as well as an SVC with $\lambda=10$ on simulated shifts: these are obtained by superposing artificial deviations on top of new series generated from the IC residuals by the BB.  Three typical deviations are simulated for each series: simple jumps ($\hat \epsilon_{\hat \mu_2(i,t)}+\delta$), trends ($\hat \epsilon_{\hat \mu_2(i,t)}+\frac{\delta}{150}(t)^{1.5}$) and oscillating shifts ($\hat \epsilon_{\hat \mu_2(i,t)} + \sin{(\eta \pi t)} \delta$).
The length of the input vector is also selected at $m=25$ (which corresponds to the 70th percentile of the run length distribution for a shift size equal to $\delta_{min}$) and $\eta=1/m$, to identify fast-oscillating shifts.
 Therefore, the SVR is allowed to predict the magnitude of shifts equal to or greater than $\delta_{min}=1.5$ in a continuous range, while the SVC discriminates between the three classes of aforementioned deviations.
The trends are implemented at time $t=0$ whereas the jumps and the oscillating shifts, quickly identified by the chart, are generated later at time $t=m$. The actual monitoring is then started at time $t \in [m,2m]$ to train the method at identifying shifts occurring at any time inside the window. 
The SVR and SVC are trained on 80\% (50400) of the instances, while the remaining 20\% (12600) are used for the validation. As explained in Appendix~\ref{sec:AC2}, the SVR shows a mean absolute percentage error (MAPE) of 26 and a normalized root mean squared error (NRMSE) of 0.26, whereas the SVC has an accuracy of 86\% on the validation set.  These values are satisfactory since the training set is relatively small with respect to the variety of deviations that are simulated. In Appendix~\ref{sec:AC3}, the confusion matrix of the SVC shows that the most frequent mis-classification, which only represents 3.9\% of all cases, is a jump predicted as an oscillating shift. This is expected since a series a jumps occurring on top of data containing disparities (see Section \ref{sec:Ia}) may look similar to an oscillating shift.  The accuracy of the SVM methods can also be further improved, by simulating a wider range of deviations (such as different power-law functions, sinusoids with varying frequencies, etc).

\subsection{Results} 
\label{sec:results}

\begin{figure}[!htb]
	\centering
	\begin{subfigure}{0.44\textwidth}
		\centering
		\includegraphics[scale=0.41]{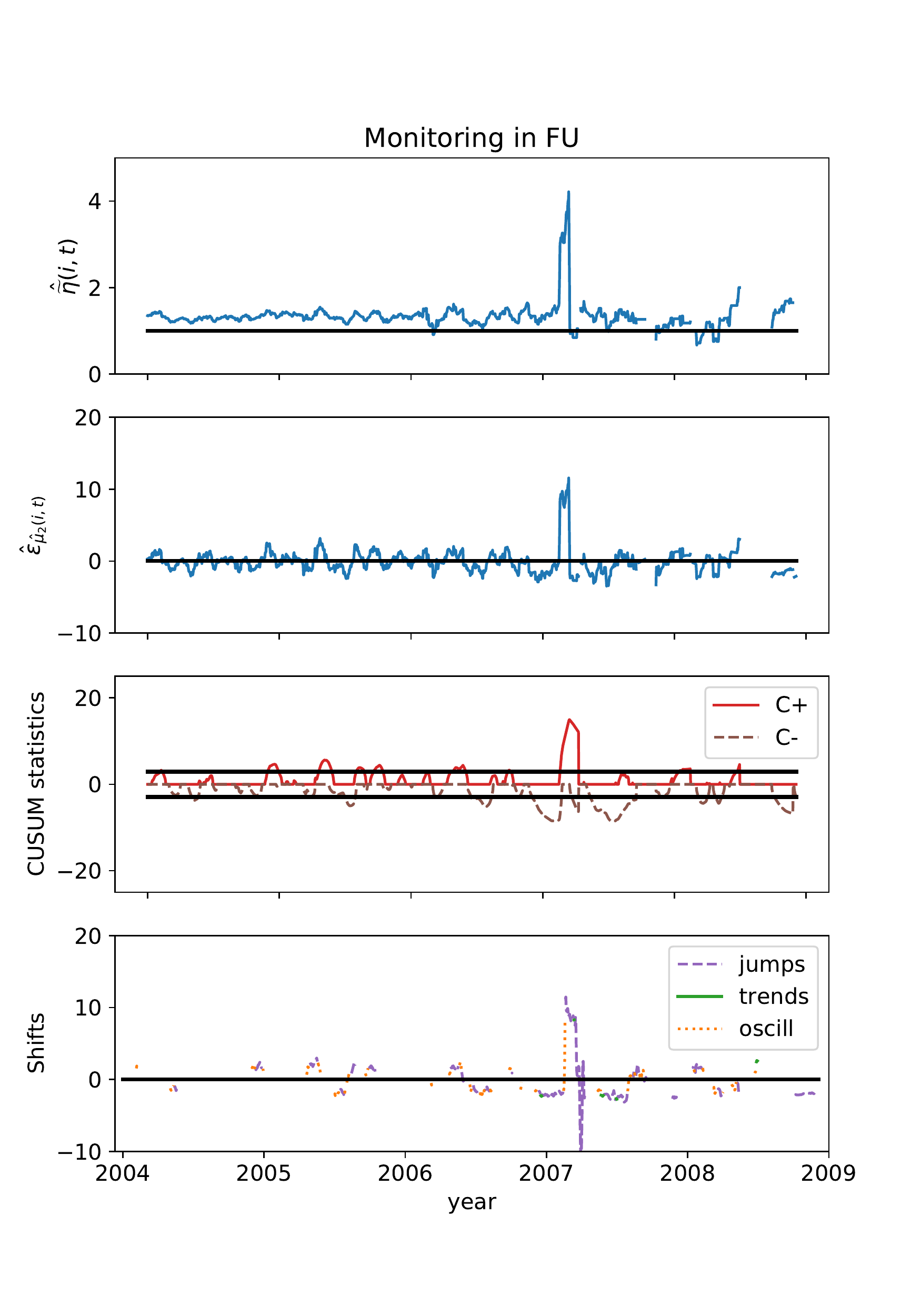}
		\caption{}
		\label{fig:FU_2008}
	\end{subfigure}
	\begin{subfigure}{0.44\textwidth}
		\centering
		\includegraphics[scale=0.41]{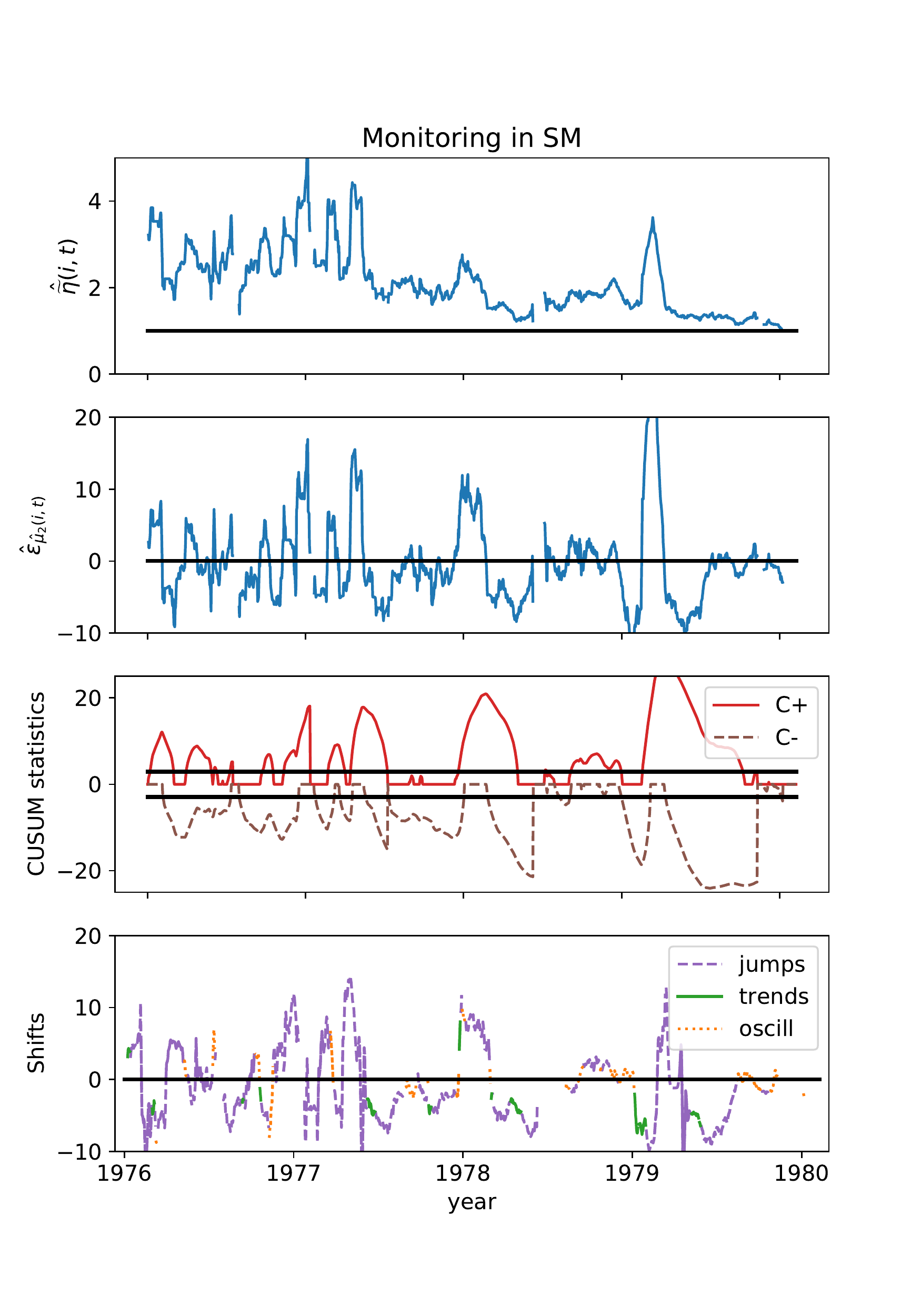}
		\caption{}
		\label{fig:SM_1979}
	\end{subfigure}
\caption{\footnotesize{ (a) Upper panel: $\hat{ \widetilde{\eta}}(i,t)$ of (\ref{E:error}) from observer Fujimori-san (FU) in Japan over the years 2004-2008. Second panel: the residuals $\hat \epsilon_{\hat \mu_2(i,t)}$ in FU \emph{with} their deviations (in addition to their disparities). Third panel: the (two-sided) CUSUM chart statistics applied on the residuals in \emph{square-root scale}. The control limits of the chart are represented by the two horizontal thick lines. For an IC average run length of $ARL_0=200$ and $k=\delta_{min}/2$, they are selected respectively at $8.5$ and $-8.5$.  Lower panel: the characteristics of the deviations predicted by the SVR and SVC after each alert. (b) Similar figure for the observatory of San Miguel (SM) in Argentina over 1976-1979. }}
\label{fig:results}
\end{figure}

Figure \ref{fig:results} shows the residuals, the CUSUM statistics and the predicted characteristics of the deviations in the observations from Fujimori-san (FU) and from the observatory of San Miguel (SM). FU is a dedicated Japanese single observer whose records are included in $P_2$, 
while SM is a severely OC station composed of different alternating observers. 
 As can be seen in the figures, the monitoring and the SVM procedures can cope with a large variety of shifts ranging from small and persistent deviations to large oscillating shifts.
In particular, the deviation reported in 2007 in FU is well identified in Figure \ref{fig:FU_2008} \citep{slides_frederic} as well as the large variations occurring in SM after 1977 in Figure \ref{fig:SM_1979} \citep[Figure~12]{Mathieu2019}. 
More identified prominent deviations as well as results for all stations are shown in Appendix \ref{sec:AD}. In addition, the chart is also able to quickly detect many other shifts, typically smaller, that are otherwise difficult to identify. 
Despite the fact that the training sets are simulated, the prediction of the SVR seems to correspond well to the pace of the residuals. The classification results can be rich. They require careful analysis to be fully explored, which opens new research perspectives. Further investigations are for instance required to see if the trends may be associated to the ageing of instruments, or if the oscillations may reflect the periodic alternation of observers in large stations. 
By combining the information about the magnitude, the form as well as the duration of the shifts, it will also be possible to send detailed reports about the statistics of the shifts back to the observers, to prevent that future major deviations build up. 
Note that the plots represent past observations, which have not been monitored by any control scheme. Consequently, the stations may stay in alert for long consecutive periods, which leads to large values of the CUSUM statistics in Figure \ref{fig:SM_1979}. 
If this approach is applied in manufacturing processes, any deviation will be promptly corrected. Therefore we expect better results for data that have already been monitored.

\section{Conclusion}
\label{sec:conclusion}

In this paper, we propose a comprehensive nonparametric approach to monitor a panel of processes, in a \emph{multiplicative} framework, experiencing a high rate of deviations along time. The procedure integrates four different techniques: control chart, block bootstrap, support vector regression and classification, 
to automatically detect and predict the characteristics of the deviations. Due to the panel structure of the data, the method does not require prior information about the observations to fully operate. 
Up to now, the method has been tested on a complex dataset: the number of sunspots ($N_s$), which is a component of $S_n$. 
In a close future, the work will be expanded not only to $N_s$ but also to the other components, namely the number of groups ($N_g$) and a composite ($N_g+10N_s$) at the basis of the $S_n$, with the objective to establish a fully automated online monitoring.  This will serve to send appropriate alerts to the stations when relevant deviations are detected. 
The scheme will enhance the future quality of the observations and the $S_n$, which is derived on those.

An additional benefit of the methodology is that the levels of the individuals processes can also be estimated and reported (after choosing appropriate time-scales to remove short-term variations). 
This will provide useful indications about the long-term drifts of the stations, as can be seen in the upper panel of Figure \ref{fig:SM_1979}.
Furthermore, the method will be extended to monitor the variances of the processes. 
We believe that the methodology can also be applied to many other streams of panel data as can be observed in manufacturing or financial data.

\bibliographystyle{apalike}
\bibliography{Submission}

\newpage 
\appendix

\section{Design of the chart}
\label{sec:AA}

The performances of the chart are usually measured using the concept of the average run length (ARL). 
The IC ARL, denoted by $ARL_0$, is the mean value of the number of samples collected from the beginning of the process to the occurrence of a false alert. It represents the rate of false positives of the chart (it is similar to the concept of type I error in hypothesis testing context). 
The OC ARL, denoted $ARL_1$, corresponds to the mean value of the number of samples collected from the appearance of a shift to the alert of the chart. It embodies the detection power of the chart (similar to the concept of type II error in hypothesis testing context). 
In practice, it is hardly possible to design a chart with $ARL_0$ as large as possible while maintaining at the same time a small value of $ARL_1$. Therefore, as in hypothesis testing, the parameters of the chart are tailored to reach the maximal detection power for a fixed rate of false positives. \\

The CUSUM chart may be designed by a bootstrap procedure for generic non-normal data, in absence of any information about their distribution. This procedure is explained as follows. 
The allowance parameter, $k$, is specified in advance to $k=\delta/2$~\citep{Moustakides1986}. Then, the control limits of the chart are adjusted by a searching algorithm that is well-described in the section 4.2.2 of~\cite{Qiu2013} for i.i.d. normal data. It is also represented specifically for the bootstrap procedure in Algorithm \ref{PA:searching}.
 From initial values of the control limit, the actual $ARL_0$ is computed on B processes that are sampled with repetition from $P_1$, the pool of $N_{IC}$ standardized IC processes. 
If the actual $ARL_0$ is inferior (resp. superior) to the pre-specified $ARL_0$, the control limit of the chart is increased (resp. decreased).
This algorithm is iterated until the actual $ARL_0$ reaches the pre-specified $ARL_0$ at the desired accuracy. \\
For the sake of clarity, this algorithm is presented for a one-side upward CUSUM chart. 
If the data are uncorrelated, the simple bootstrap may be used to resample the observations of $P_1$. 
Otherwise, the observations are sampled per blocks of consecutive values (block bootstrap method). This distinction is denoted respectively by 'per single observation' or 'per blocks' in the algorithm.  \\
We make the following comments about the design previously explained. 
\begin{enumerate}
\item The choice of an appropriate $\delta$ depends on the SPC problem.  In absence of information about the deviations, a proper value of $\delta$ may be estimated from the OC processes by a recursive method described in the following subsection \ref{sec:AA1}.
\item The second remark concerns the value of $k$. The relation $k=\delta/2$ can be suboptimal if the distribution of the IC residuals differs significantly from the normal. In this case, $k$ may be selected numerically for a particular value of $\delta$, as described in subsection \ref{sec:AA2}.
\end{enumerate}

\begin{algorithm}[H]
\DontPrintSemicolon
\LinesNotNumbered 
\tcc{Select the control limit of the chart using the bootstrap method}
Select values for: \;
[$h_U$,$h_L$], the interval where $h$ is searched on \;
$ARL_0$, the desired value of the in-control ARL \;
$\rho$, the accuracy to reach $ARL_0$ \;
$\delta$, the typical shift size and set $k=\delta/2$ \;
\;
\While{$|ARL-ARL_0|>\rho$}{
$h=\frac{h_U+h_L}{2}$ \;
\For{b in (1:B)}{
\tcp{sample data with repetitions from the $N_{IC}$ (IC) processes:}
resample($\hat \epsilon_{\hat \eta}(i_{IC},t)$) per single observation or per blocks \; 
run the control chart $C$ on the resampled data \;
\uIf{the chart gives an alert ($C>h$)}{
$RL[b]$ $=$ time of alert \; }
\Else{
$RL[b]$ $=$ large number \;
}
}
$ARL$ = mean($RL$) \;
$h_U=(h_L+h_U)/2$, $h_L=h_L$ if $ARL_0 > ARL$ \;
$h_L=(h_L+h_U)/2$, $h_U=h_U$ if $ARL_0 < ARL$ \;
}
$h=\frac{h_U+h_L}{2}$ \;
\;
 \caption{Searching algorithm for the control limit}
\label{PA:searching}
\end{algorithm}
\noindent
\vspace{1.45cm}

\subsection{Selection of an appropriate value for the shift size}
\label{sec:AA1}

In absence of information about the size of the shifts, an appropriate value of $\delta$ may be estimated using a recursive method described in Algorithm \ref{PA:size}.
From an initial value of $\delta$, denotes by $\delta_0$, we compute the control limit of the chart for $k=\delta_0/2$ as explained in Algorithm \ref{PA:searching}. Then, this chart is applied on the bootstrapped OC residuals to estimate the size of the actual shifts that the processes experience along time.
For each Monte-Carlo simulation, the magnitude of a deviation is computed after the alert with the following formula from~\cite{Montgomery2005}:
\begin{equation}
\label{E:Montgomery}
\hat \delta =    \begin{cases}
\begin{tabular}{cl}
$k+\frac{C_i^+}{N^+}$ & if $C_i^+>h$  \\
$-k-\frac{C_i^-}{N^-}$ & if  $C_i^-<-h$,
\end{tabular}
    \end{cases}
\end{equation}
where $\hat \delta$ is the estimated shift size expressed in standard deviation units and $N^+$ (resp. $N^-$) represents the number of observations where the CUSUM statistics $C_i^+$ (resp. $C_i^-$) has been non-zero. 
Although naturally valid for uncorrelated and normally distributed observations, the above formula can still be used in general to provide a rough approximation of the shifts. 
Then, a new value for $\delta$ is computed as a particular quantile of the distribution of the OC shifts, which depends on the SPC problem. Indeed, according to the situation, it might be interesting to detect for instance either the smallest or the most common shift.
The algorithm is iterated few times until convergence.

\begin{algorithm}[H]
\DontPrintSemicolon
\LinesNotNumbered 
\tcc{Estimate an appropriate shift size using the bootstrap method}
Select values for: \;
$\delta_0$, an initial value of $\delta$ \;
$\rho$, the accuracy of the algorithm's convergence \; 
\;
\While{$|\delta_0-\delta|>\rho$}{
Compute the control limit $h$ using algorithm 1 for $k=\delta_0/2$ \;
\For{b in (1:B)}{
\tcp{sample data with repetitions from the OC processes:}
resample($\hat \epsilon_{\hat \eta}(i_{OC},t)$) per single observation or per blocks \; 
run the CUSUM chart $C$ on the resampled data \;
\uIf{the chart gives an alert ($C>h$)}{
$shift\_size[b]$ $=$ $k + \frac{C}{N}$ (Montgomery's formula)\; }
}
$\delta$ = $quantile$($shift\_size$) \;
$\delta_0=\delta$ \;
}
$\delta$ 
\;
 \caption{Algorithm to estimate the size of the actual deviations}
\label{PA:size}
\end{algorithm}
\noindent
\vspace{1.45cm}

\subsection{Selection of an optimal value for the allowance parameter}
\label{sec:AA2}

If the distribution of the IC residuals differs significantly from the normal, $k$ may be selected numerically for a particular value of $\delta$ as described in Algorithm \ref{PA:threshold}. The control chart is first adjusted for different values of $k$ around $k=\delta/2$ to reach a common pre-specified value of $ARL_0$. For each pair ($k$,$h$), the actual detection power of the chart is evaluated using simulated shifts (i.e. bootstrapped IC residuals perturbed by a jump of size $\delta$). The value of the pair ($k$,$h$) yielding the best performance (i.e. the smallest $ARL_1$) is then selected for the actual design of the chart. \\

\begin{algorithm}[H]
\DontPrintSemicolon
\LinesNotNumbered 
\tcc{Compute an optimal value of $k$ for a particular $\delta$ }
Select values for: \;
$\delta$, the shift size (algorithm 2 can be used) \;
$K$, a set of values for $k$ around $k=\delta/2$ \; 
\;
\For{k in (K)}{
Compute the control limit $h$ using algorithm 1 for $k$ \;
\For{b in (1:B)}{
\tcp{sample data with repetitions from the IC processes:}
data $=$ resample($\hat \epsilon_{\hat \eta}(i_{IC},t)$) per single observation or per blocks \; 
\tcp{add a simulated shift of size $\delta$ on the resampled data} 
data $+$ $\delta$ \;
run the CUSUM chart $C$ on the resampled data \;
\uIf{the chart gives an alert ($C>h$)}{
$RL_1[b]$ $=$ time of alert \; }
\Else{
$RL_1[b]$ $=$ large number \; }
}
$ARL_1[k]$ = mean($RL_1$) \;
}
$k$ = $K[argmin(ARL_1)]$ \;
\;
 \caption{Algorithm to optimize the value of the allowance parameter}
\label{PA:threshold}
\end{algorithm}
\noindent
\vspace{1.45cm} 

\section{Bootstrap vs parametric approach}
\label{sec:AB}

In this section, we compare the performances of two CUSUM charts, one designed by the block bootstrap (BB) method and the other adjusted using a parametric approach, on non-trivially autocorrelated data. 
To this end, series with a complex autocorrelation structure are first simulated. Positively-autocorrelated series are created (arbitrarily) by adding a small positive trend on data correlated by an ARMA(3,3). Negatively-autocorrelated series are also generated by adding a short-term seasonality on top of the trend. 
A sample of 40 time series with 500 observations are generated from those positively and negatively autocorrelated models. 
 Thereafter, in the classical approach, each series is decorrelated by an ARMA(2,2) (similar results are obtained with an ARMA(3,3)) and a non-central student ($nct$) distribution~\citep{distributions} is fitted on the distribution of the residuals, assumed to be decorrelated. 
 The control limits are then computed either by the BB procedure as explained in Algorithm \ref{PA:searching} or classically using the $nct$ distribution of the ARMA residuals. Finally, new time series are generated from the models and the \emph{actual} $ARL_0$s are evaluated for the BB and the parametric approach. \\

\begin{figure}[htbp]
	\centering
		\includegraphics[scale=0.7]{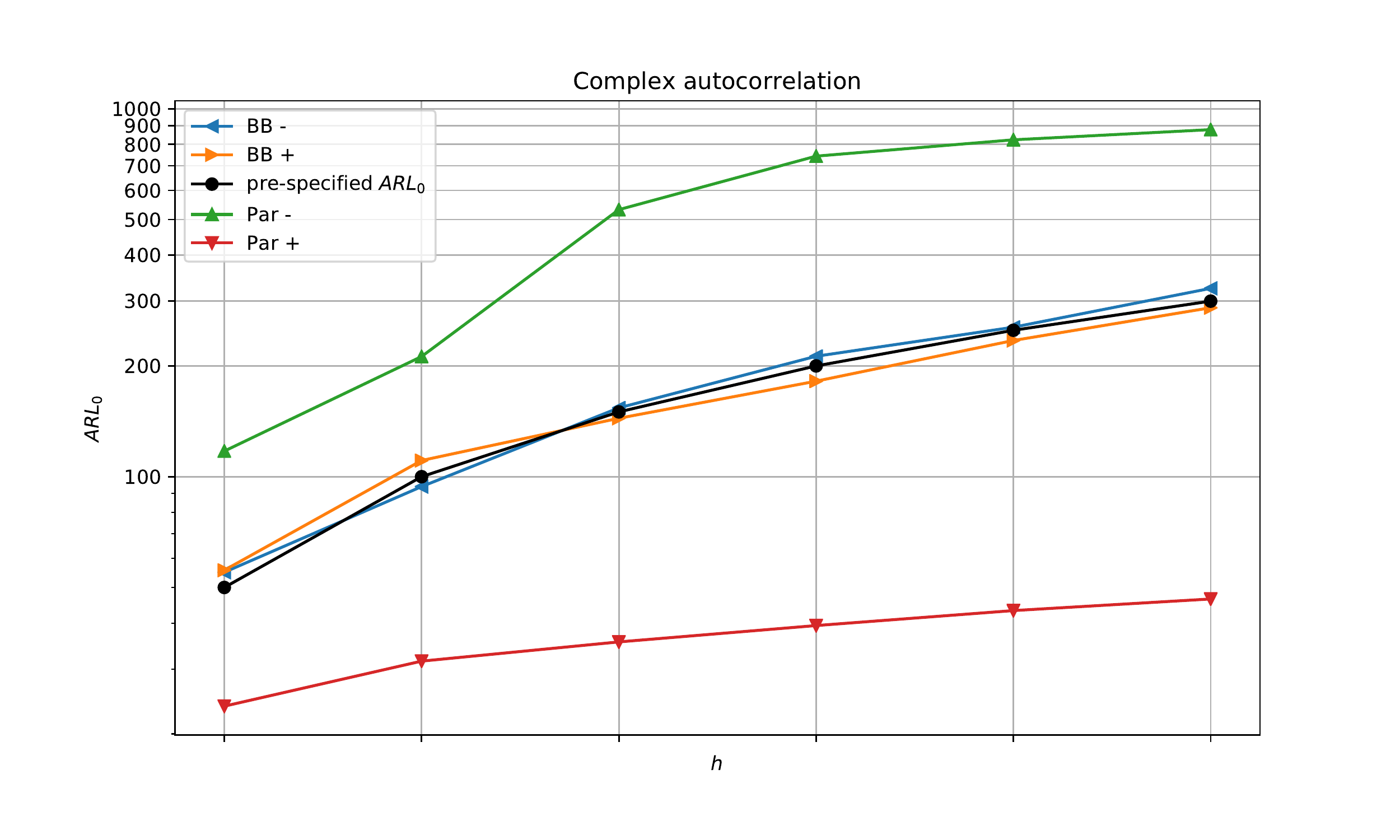}
		\caption{\footnotesize{$ARL_0$ values of the two-sided CUSUM chart designed by the BB (``BB'' in the legend) with a block length equal to 50 and by a parametric approach (denoted by `` par'' in the legend). The target values for $ARL_0$ are represented by ``pre-specified $ARL_0$''. The chart is designed on data experiencing a non-trivial positive (resp. negative) autocorrelation, which is represented by the symbol `` $+$'' (resp. ``$-$'') in the legend. The number of MC replications is set to 4000 and $k=0.75$. The x-axis corresponds to the values of the control limits, which differ for all curves. To ensure a better readability, the x-axis is thus represented without values (which allows us to vertically align the curves) while the y-axis is on a logarithmic scale.  }}
	\label{fig:BB_param}
\end{figure}

\begin{figure}[hbtp]
	\centering
		\includegraphics[scale=0.7]{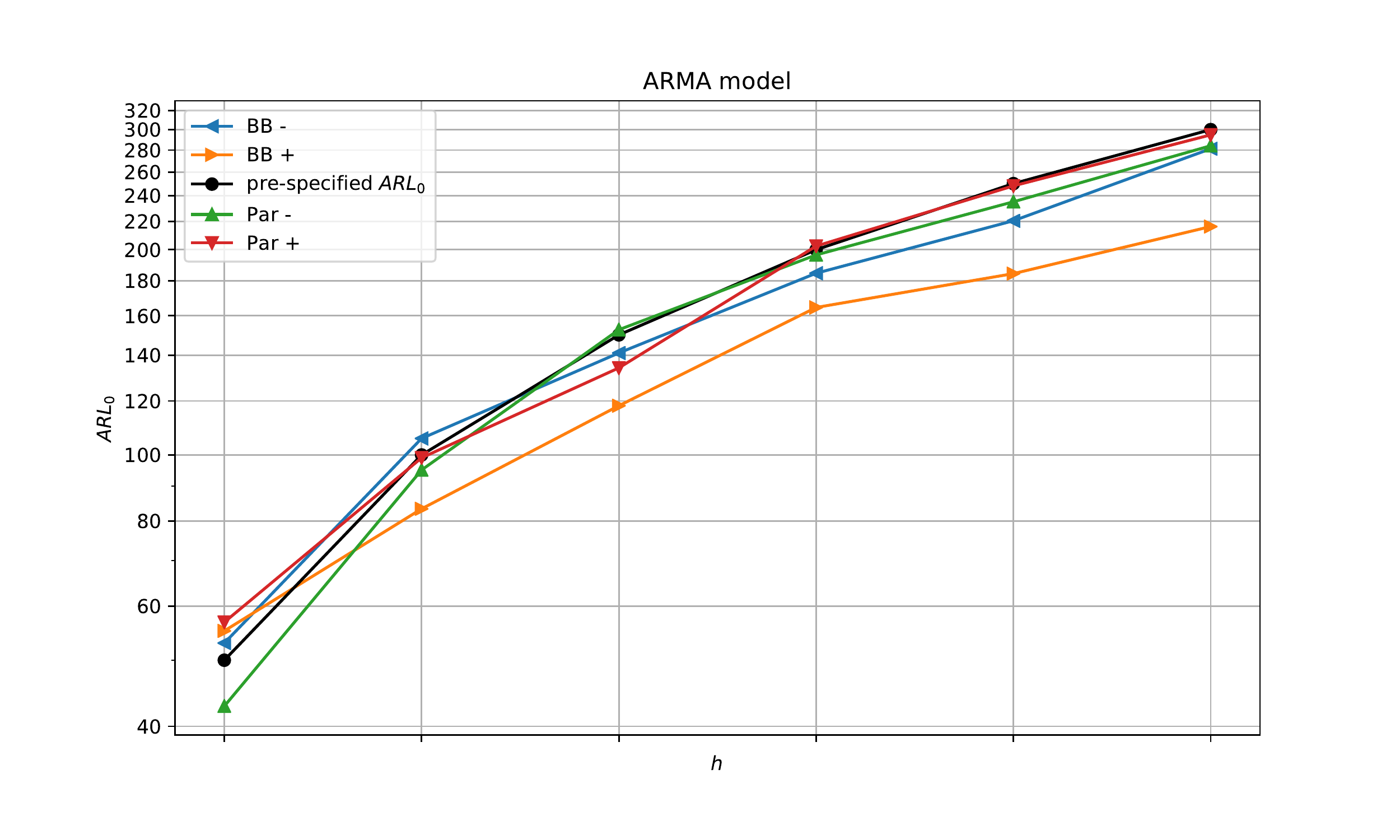}
		\caption{\footnotesize{$ARL_0$ values of the two-sided CUSUM chart designed by the BB (``BB'' in the legend) with a block length equal to 50 and by a parametric approach (denoted by `` par'' in the legend).  The target values for $ARL_0$ are represented by ``pre-specified $ARL_0$''.  The chart is designed on data correlated by an ARMA(1,1) model with positive (resp. negative) autoregressive coefficient, which is represented by the symbol `` $+$'' (resp. ``$-$'') in the legend. The number of MC replications is set to 4000 and $k=0.75$. The x-axis corresponds to the values of the control limits, which differ for all curves. To ensure a better readability, the x-axis is thus represented without values (which allows us to vertically align the curves) while the y-axis is on a logarithmic scale.  }}
	\label{fig:BB_param_2}
\end{figure}

As can be seen in Figure \ref{fig:BB_param}, the actual $ARL_0$ values of the chart are significantly higher (resp. smaller) than the pre-specified $ARL_0$s when the data are negatively (resp. positively) autocorrelated in the parametric approach. These differences arise since the autocorrelation of the initial series is complex and not-well represented by a simple ARMA model. Therefore, the residuals still contain some autocorrelation that is neglected when adjusting the control limits. 
On the contrary, in the BB, the control limits are better adjusted, since this method takes into account the autocorrelation of the data. \\
In the following, we repeat the procedure with data generated from an ARMA(1,1) model. The parameters of the model are selected at $\phi=0.8$ (AR), $\theta=0.2$ (MA) and $\phi=-0.8$, $\theta=0.2$ respectively, for the positively and negatively autocorrelated models.
As can be seen in Figure \ref{fig:BB_param_2}, the parametric approach performs better here than the bootstrap, as expected. 
The actual $ARL_0$s of the BB are however close to those of the parametric approach. 
Another simulation, not presented here, shows that by increasing the block length, the performance of the BB can still be improved. 

\FloatBarrier

\section{Support vector machine}
\label{sec:AC}

The support vector machine (SVM)~\citep{Vapnik1998} is a supervised machine-learning procedure used for classification and regression. 
In the following, we briefly introduce the support vector regression (SVR) since it is similar but less well-known than the SVM classifier (SVC). Readers may also refer to~\cite{Burges1998} for a generic description of the SVC and to ~\cite{Smola2004} for more detailed explanations about SVR. \\
We denote by \{ $\bm{x}^j, y^j| j=1,2, M$ \} the $M$ training pairs where $x \in \mathcal{R}^m$ represents a training example  and $y \in \mathcal{R}$ is the label.  
A training example typically corresponds to a series of IC data artificially corrupted by a particular shift function ($\epsilon_{\hat \eta}(i_{ic},t)+f(\delta; i,t)$), whose corresponding label is the magnitude of the shift ($\delta$).
Mathematically, the SVR regression function writes as:
\begin{equation}
f(\bm{x})=\bm{w}^T \phi(\bm{x})+b.
\end{equation}
where $\phi$ is a non-linear function mapping the input data into a higher dimensional feature space, where the non-linear regression problem may be expressed into a simpler linear regression. 
The training aims at computing the coefficients $\bm{w}$ and $b$, by solving the following optimization problem: 
\begin{equation}
\label{E:regloss}
\underset{\bm{w},b}{\text{min}} \frac{1}{2}||\bm{w}||^2 + \lambda \frac{1}{M} \sum_{j=1}^M L_{\epsilon}(y^j, f(\bm{x})),
\end{equation}
where
\begin{equation*}
L_{\epsilon}(y^j, f(\bm{x})) =  
\begin{cases}
\begin{tabular}{cl}
$|y-f(\bm{x})|-\epsilon$ & $\quad |y-f(\bm{x})| \geq \epsilon$   \\
$0$ & otherwise.
\end{tabular}
 \end{cases}
\end{equation*}
The parameter $\lambda$ represents a trade-off between misclassification and regularization and $\epsilon$ is equivalent to an approximation accuracy, i.e. errors below $\epsilon$ are neglected.
This optimization problem may be rewritten with Lagrange multipliers as a dual problem and may be easily solved thanks to the introduction of a kernel function $K(\bm{x^j}\bm{x^k})=\phi(\bm{x^j})\phi(\bm{x^k})$. 
This step, commonly referred to as the  ``kernel trick'' in the literature, ensures that the dual problem may be solved without the explicit form of the mapping $\phi$. Hence, all computations can be performed in the input space, which saves much computational time. Consequently, the choice of the kernel function is an important hyper-parameter of the method that should be carefully selected. 
Another substantial parameter of the SVM procedures is the dimension $m$ of the instances $\bm{x}$, which corresponds to the length of the input vector that is fed into the SVR (or SVC). Different methods for choosing an appropriate value for $m$ are described in the following. 

\subsection{Size of the input vector}
\label{sec:AC1}

The length $m$ of the input vector  may be numerically calculated to obtain the best accuracy while maintaining the computing efficiency of the method. Alternatively, $m$ may also be selected by a more data-driven approach described in Algorithm \ref{PA:m}. This method is based on the distribution of the OC run length and may be explained as follows. 
We recall that we denote by $\delta_{min}$ the size of the smallest shift that we aim to detect.
$\delta_{min}$ may be chosen using a priori information about the processes. Alternatively, if such information are unavailable, it may be estimated using the Algorithm \ref{PA:size}. 
Then, we simulate IC observations shifted by $\delta_{min}$ ($\epsilon_{\hat \eta}(i_{ic},t)+\delta_{min}$) using the BB and compute the run lengths of the chart. 
The length of the input vector may finally be selected as the 90-th quantile or another appropriate upper quantile of the run length distribution. 
The aim is to choose $m$ sufficiently large to ensure that most of the time the smallest shifts of size $\delta_{min}$ (which are the hardest to detect) will be recognized, while maintaining the computing efficiency of the method.
Note that, if the processes undergo oscillating deviations, the window length should not be longer than half a period of the fastest oscillations. Otherwise, these shifts may not be correctly identified by the SVR (or SVC).

\begin{algorithm}[H]
\DontPrintSemicolon
\LinesNotNumbered 
\tcc{Estimate an optimal value for $m$, the length of the input vector }
Select value for: \;
$\delta_{min}$, the smallest shift size that we aim to detect (algorithm 2 can be used) \;
\;
Compute the control limit $h$ using algorithm 1 for $k=\delta_{min}/2$ \;
\For{b in (1:B)}{
\tcp{sample data with repetitions from the IC processes:}
data $=$ resample($\hat \epsilon_{\hat \eta}(i_{IC},t)$) per single observation or per blocks \; 
\tcp{add a simulated shift of size $\delta_{min}$ on the resampled data} 
data $+$ $\delta_{min}$ \;
run the CUSUM chart $C$ on the resampled data \;
\uIf{the chart gives an alert ($C>h$)}{
$RL_1[b]$ $=$ time of alert \; }
\Else{
$RL_1[b]$ $=$ large number \; }
}
$m$ = quantile($RL_1$) \;
\;
 \caption{Algorithm to select the size of the input vector}
\label{PA:m}
\end{algorithm}
\noindent
\vspace{\baselineskip}

\subsection{Performance criteria}
\label{sec:AC2}

The prediction ability of the SVR may be evaluated on the validation set with two criteria: the mean absolute percentage error (MAPE) and the normalized root mean squared error (NRMSE). They are defined by the following expressions: 
\begin{equation}
MAPE=\frac{1}{M}\sum_{j=1}^M|\frac{y^j -\hat y^j}{y^j}| \times 100\%,
\end{equation}
and
\begin{equation}
NRMSE=\sqrt{\frac{\sum_{j=1}^M (y^j-\hat y^j)^2}{\sum_{j=1}^M (y^j)^2} },
\end{equation}
where $\hat y^j$ is the label predicted by the SVR, corresponding to the instance $\bm{x}^j$. 

The performance of SVC may also be evaluated using the classification accuracy: 
\begin{equation}
ACCURACY=\frac{1}{M}\sum_{j=1}^M \mathbbm{1}_{\{\hat y_{\Gamma}^j = y_{\Gamma}^j \}} \times 100\%,
\end{equation}
where $\hat y_{\Gamma}^j$ denotes the label predicted by the SVC among the $\Gamma$ different classes, $\hat y_{\Gamma}^j \in [1,...\Gamma]$ and  $y_{\Gamma}^j$
is the true label of the instance $\bm{x}^j$. \\
The accuracy is a measure of the total performance of the classifier, for all classes. To obtain a detailed view of the performance of the method for each class, we can compute the confusion matrix. The columns of this matrix represent the prediction labels $\hat y_{\Gamma}^j$ whereas the rows correspond to the true labels $y_{\Gamma}^j$. Hence, the elements of the matrix on the main diagonal correspond to the numbers of correct classification per class, while the other entries show the number and the type of classification errors. 

\subsection{Performances of the application}
\label{sec:AC3}

The confusion matrix of the classification problem described in Section~\ref{sec:appli_II} is written in Table~\ref{tab:cm} for the 12600 testing examples.

\begin{table}[ht]
\begin{center}
\begingroup
\begin{tabular}{lccc}
\hline
\hline
Deviation & \multicolumn{3}{c}{Predicted} \\
\cline{2-4}
True value & jump & trend & oscillating shift\\
\hline
jump & 3546 & 164 & 490 \\
trend &  305 & 3604 & 291 \\
oscillating shift & 262 & 191 & 3747 \\
\hline
\end{tabular}
\endgroup
\end{center}
\caption{\small{Confusion matrix.}}
\label{tab:cm}
\end{table}


This matrix is also rewritten in percentages in Table~\ref{tab:cm2}.

\begin{table}[ht]
\begin{center}
\begingroup
\begin{tabular}{lccc}
\hline
\hline
Deviation & \multicolumn{3}{c}{Predicted} \\
\cline{2-4}
True value & jump & trend & oscillating shift \\
\hline
jump & 28.1 & 1.3 & 3.9 \\
trend &   2.4 & 28.6 & 2.3 \\
oscillating shift & 2.0 & 1.5 & 29.7 \\
 \hline 
\end{tabular}
\endgroup
\end{center}
\caption{\small{Confusion matrix, expressed in percentages.}}
\label{tab:cm2}
\end{table}

With a perfect classifier, the matrix would be diagonal with elements 1/3 ($\approx 33.3$\%), since the classes are balanced. As can be seen for the actual classifier of Section~\ref{sec:appli_II}, the most frequent errors, representing 3.9\% of the total cases and 29\% of the mis-classifications, are jumps predicted as oscillating shifts. Those are expected since a series of consecutive jumps may look similar to an oscillating shift, especially on data containing disparities (see Section \ref{sec:Ia}).  

\section{Further results}
\label{sec:AD}

In this section, we present complementary results about the different stations. 
We first present in the next subsection some prominent deviations that have been manually identified in several observatories at different times. Then, we show in the second subsection the complete scheme applied on all stations over a selected period. 
The figures are composed of four panels: the upper panel exposes what corresponds to $\hat{ \widetilde{\eta}}$ of (\ref{E:error}) for a particular station, the second panel shows the residuals defined in (\ref{E:residuals_sun}) for the station, the third panel represents the CUSUM statistics (and the control limits of the chart) applied on the residuals in square-root scale whereas the lower panel displays the characteristics (magnitude and form) of the shifts predicted by the SVC and SVR. The control limits of the chart are selected respectively at 8.5 and -8.5 to reach an $ARL_0$ value of 200, for $k=0.75$. 
Note that the stations of the pool, $P_1$, (and thus also the stations included in $P_2$) are shown with all their deviations in addition to their \emph{disparities} (see Section \ref{sec:Ia}).

\subsection{Prominent deviations}
\begin{figure}[!htb]
	\centering
	\begin{subfigure}{0.44\textwidth}
		\centering
		\includegraphics[scale=0.41]{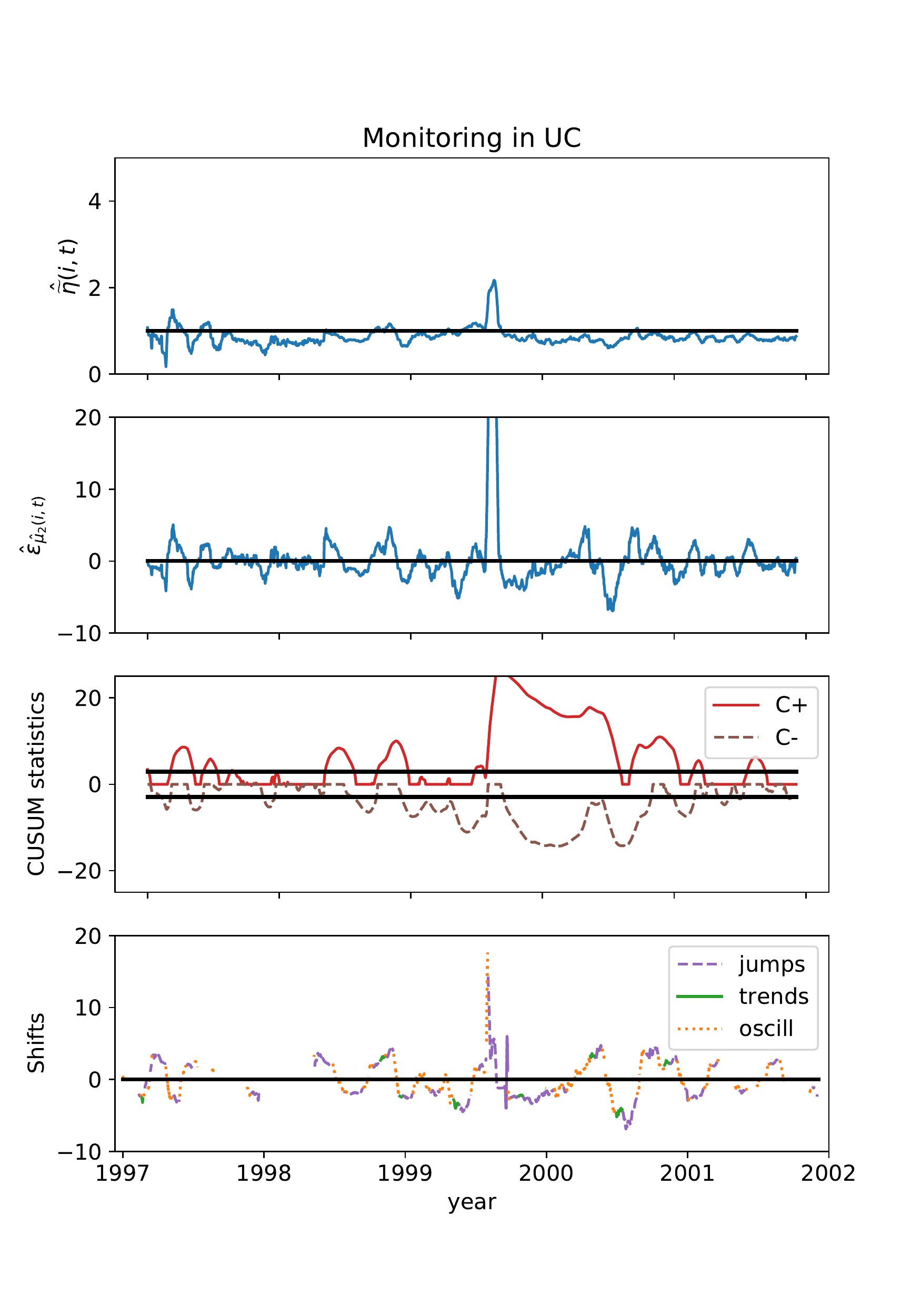}
		\caption{}
		\label{fig:UC_1999}
	\end{subfigure}
	\begin{subfigure}{0.44\textwidth}
		\centering
		\includegraphics[scale=0.41]{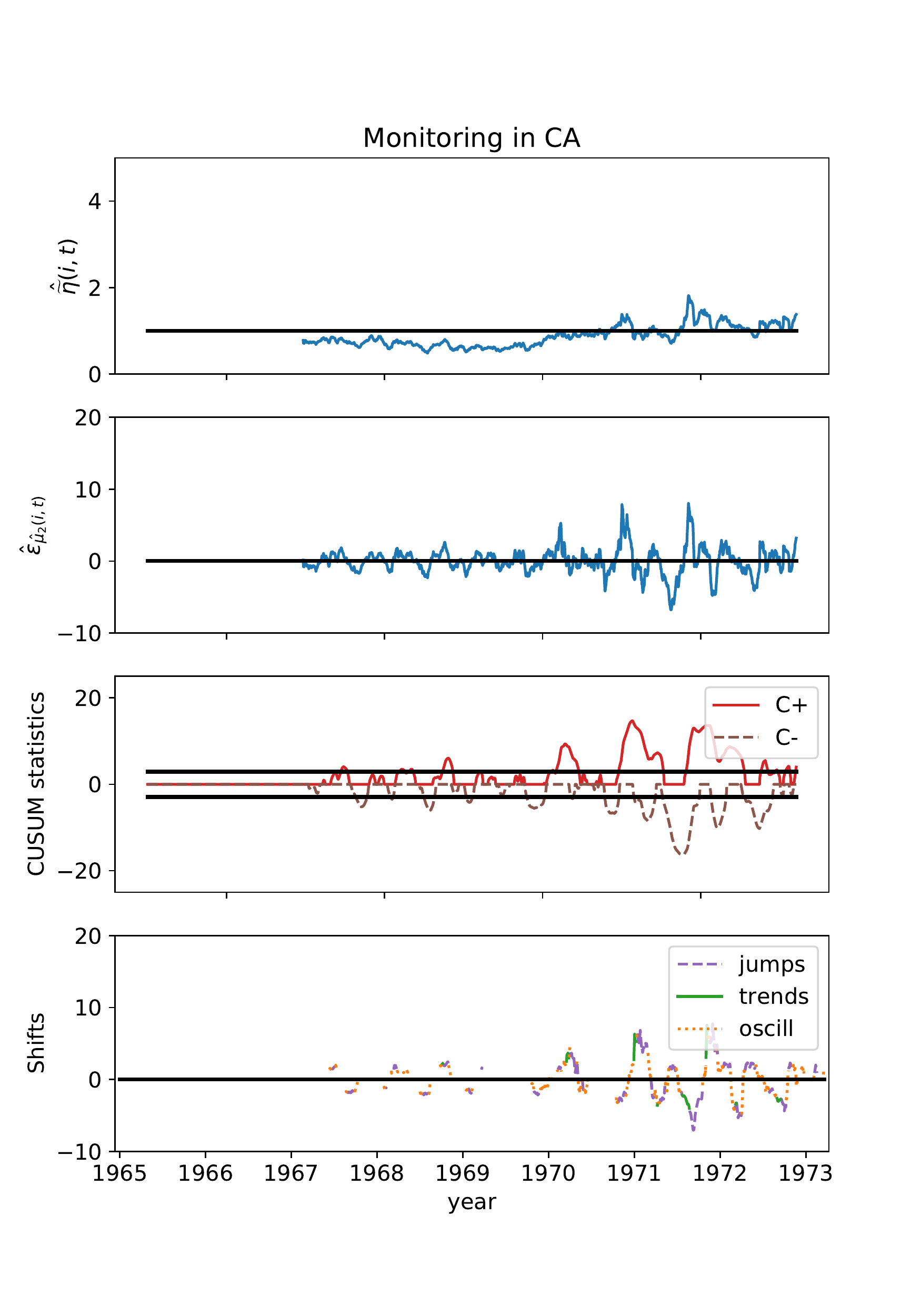}
		\caption{}
		\label{fig:CA_1970}
	\end{subfigure}
\caption{\footnotesize{ a) The control scheme applied on the data from the USET station (UC) in Belgium over the years 1997-2001.  b) Similar figure for the station Catania (CA) in Italy over 1965-1972. }}
\label{fig:dev1}
\end{figure}

\begin{figure}[!htb]
	\centering
	\begin{subfigure}{0.44\textwidth}
		\centering
		\includegraphics[scale=0.41]{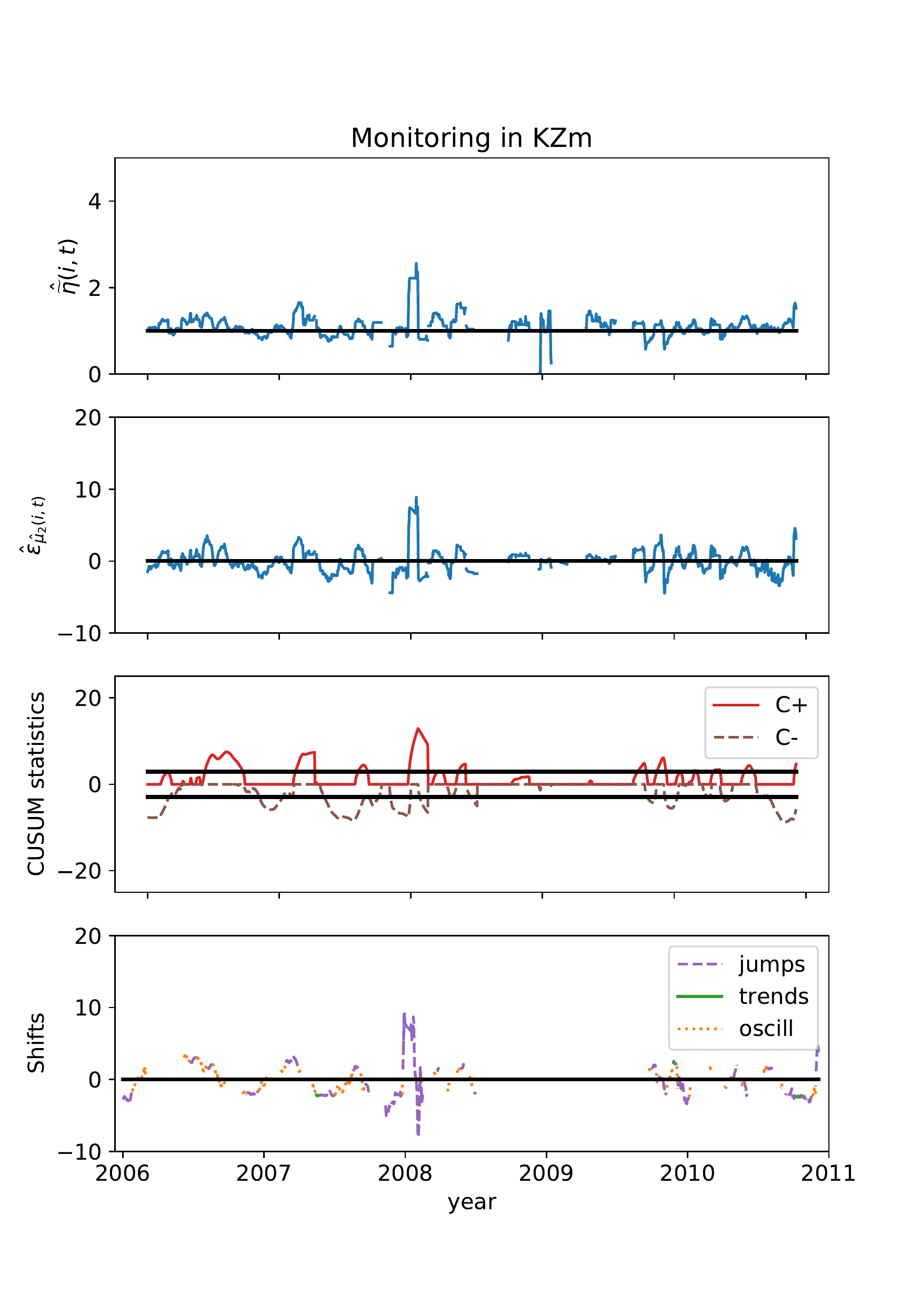}
		\caption{}
		\label{fig:KZm_2008}
	\end{subfigure}
	\begin{subfigure}{0.44\textwidth}
		\centering
		\includegraphics[scale=0.41]{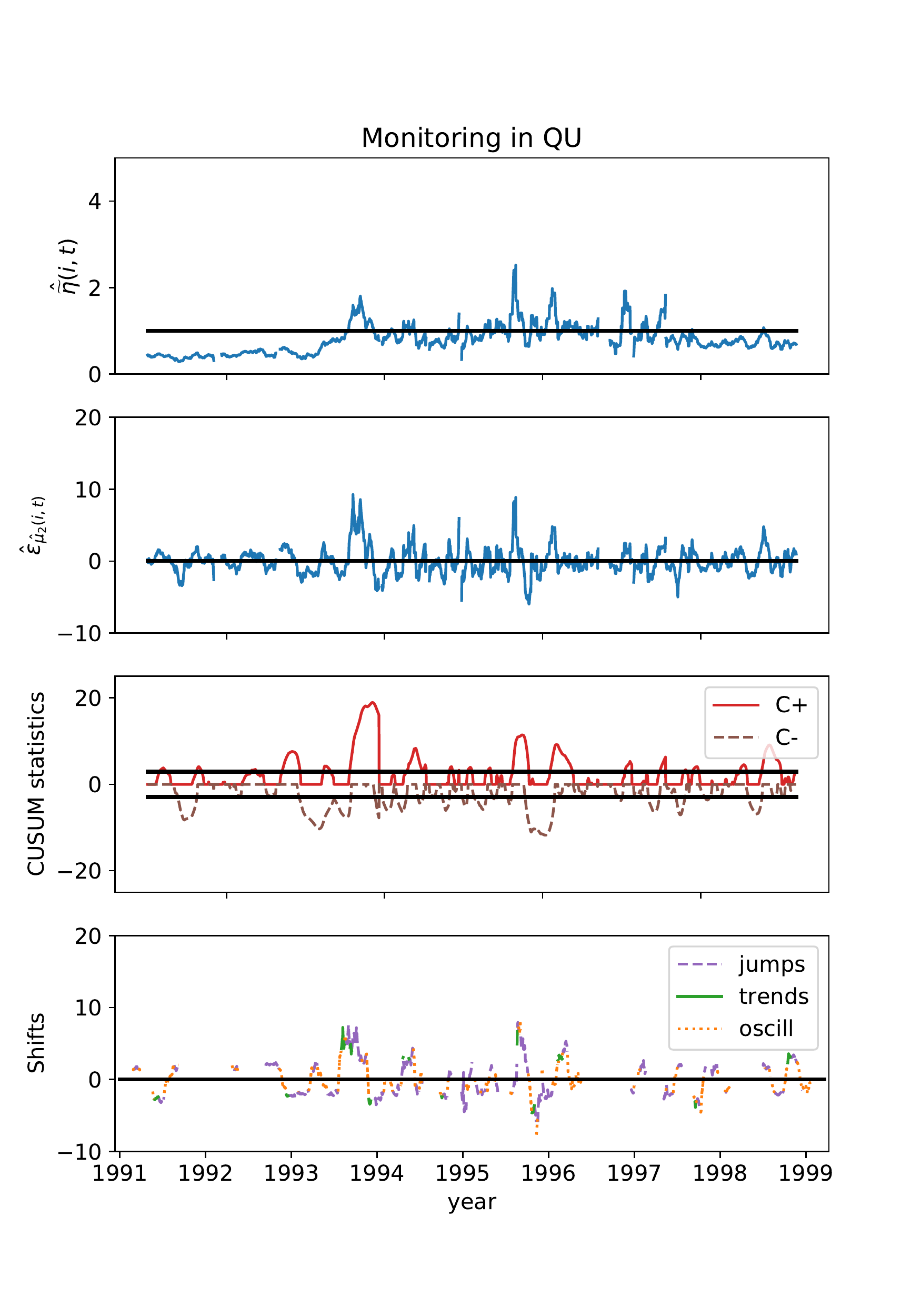}
		\caption{}
		\label{fig:QU_1994}
	\end{subfigure}
\caption{\footnotesize{ (a) The control scheme applied on the data from the station Kanzelhobe (KZm) in Austria over the years 2006-2010.  b) Similar figure for the station Quezon (QU) in Philippines over 1991-1998. }}
\label{fig:dev2}
\end{figure}

During different recalibration procedures, several important deviations have been identified in the dataset, such as a brief but large shift reported in the USET station around mid-1999.  This shift is well detected by the control scheme as can be seen in Figure \ref{fig:UC_1999} as well as a similar deviation occurring in the observatory of Kanzelhobe at the start of 2008 in Figure \ref{fig:KZm_2008}.
Important variations that extend over several years have also been identified in the Catania observatory starting from 1970 until 1973 and in Quezon from mid-1993 to mid-1996 \citep{slides_frederic}. Those are also quickly recognized by the control scheme in Figure \ref{fig:CA_1970} and \ref{fig:QU_1994}, respectively. 
This demonstrates that the control scheme is able to automatically detect large deviations that have been only recently identified manually.
Even more interesting is the large variety of shifts, otherwise difficult to discover, that the chart detects in addition to the major deviations. These shifts are typically smaller but may lead to major deviations if there are not rapidly detected and reported. \\
Note that the level of the stations (visible in the upper panel) varies with time, as can be seen for instance over 1970-1972 in Figure \ref{fig:CA_1970}. Therefore, it might also be interesting to monitor these levels, which correspond to long-term drifts, persisting over several years.  

\FloatBarrier

\subsection{Monitoring of all stations over 1985-1989}

The remaining figures show the complete scheme applied on all stations over the period 1985-1989. This period corresponds to the end of solar cycle 21 and contains most of the ascending part of cycle 22. It is also a period where all stations are observing (i.e. the data do not contain long series of missing values on this time interval). \\
Over the years 1985-1989, most of the stations do not exhibit particular deviations previously detected by hand, apart from Kandilli in Figure \ref{fig:KH} (around 1986 and 1987) and the USET station in Figure \ref{fig:UC} (large deviation after 1988, see \cite{slides_frederic}). The observatory of Locarno also shows a cluster of large variations over 1986-1988, represented in the enlarged Figure \ref{fig:LO} for readability. 
The other figures illustrate the large variety of deviations that the stations experience along time as well as the performance of our control scheme in such complex situations.

\begin{figure}[!htb]
	\centering
	\begin{subfigure}{0.44\textwidth}
		\centering
		\includegraphics[scale=0.41]{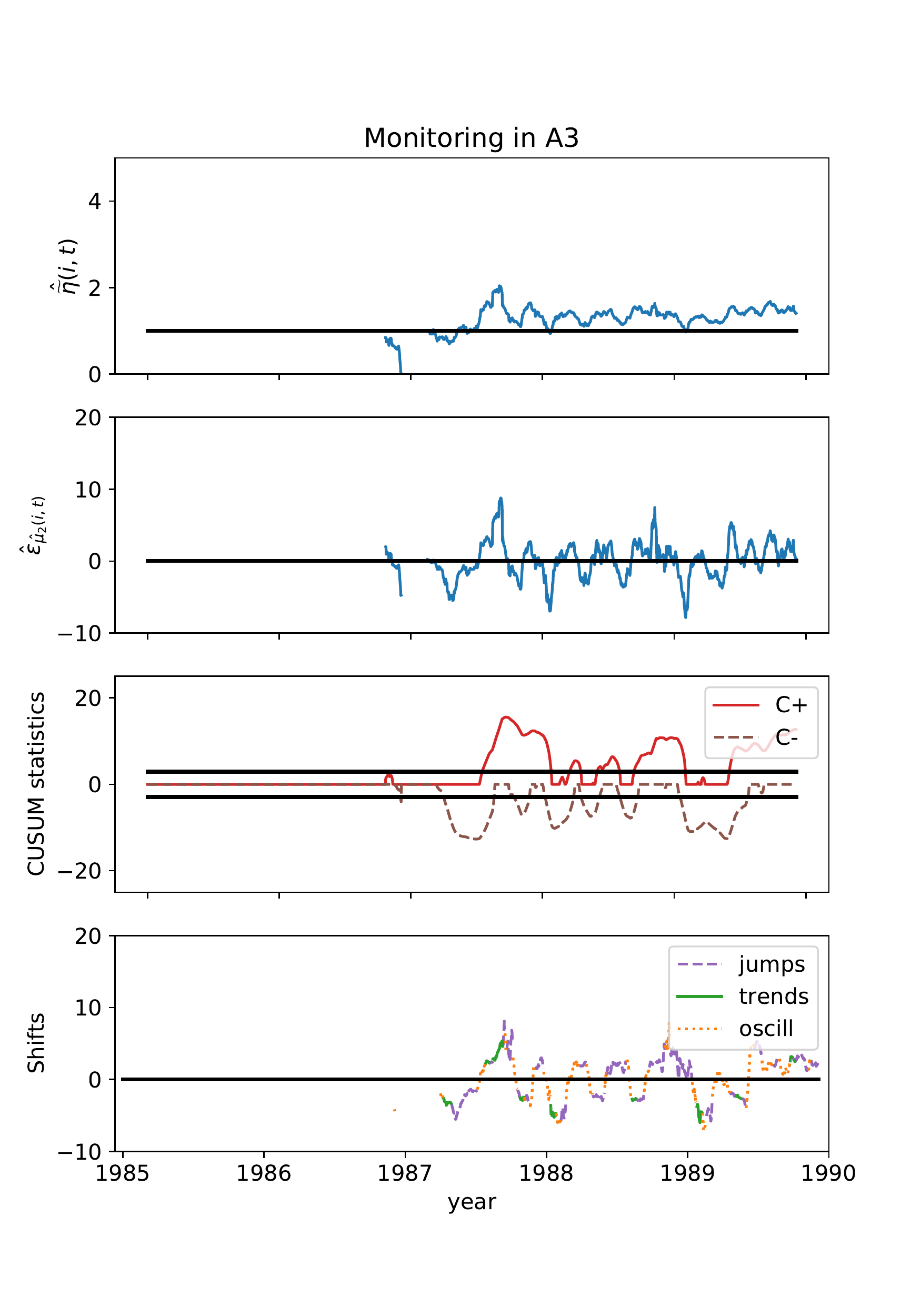}
		\caption{}
		\label{fig:A3}
	\end{subfigure}
	\begin{subfigure}{0.44\textwidth}
		\centering
		\includegraphics[scale=0.41]{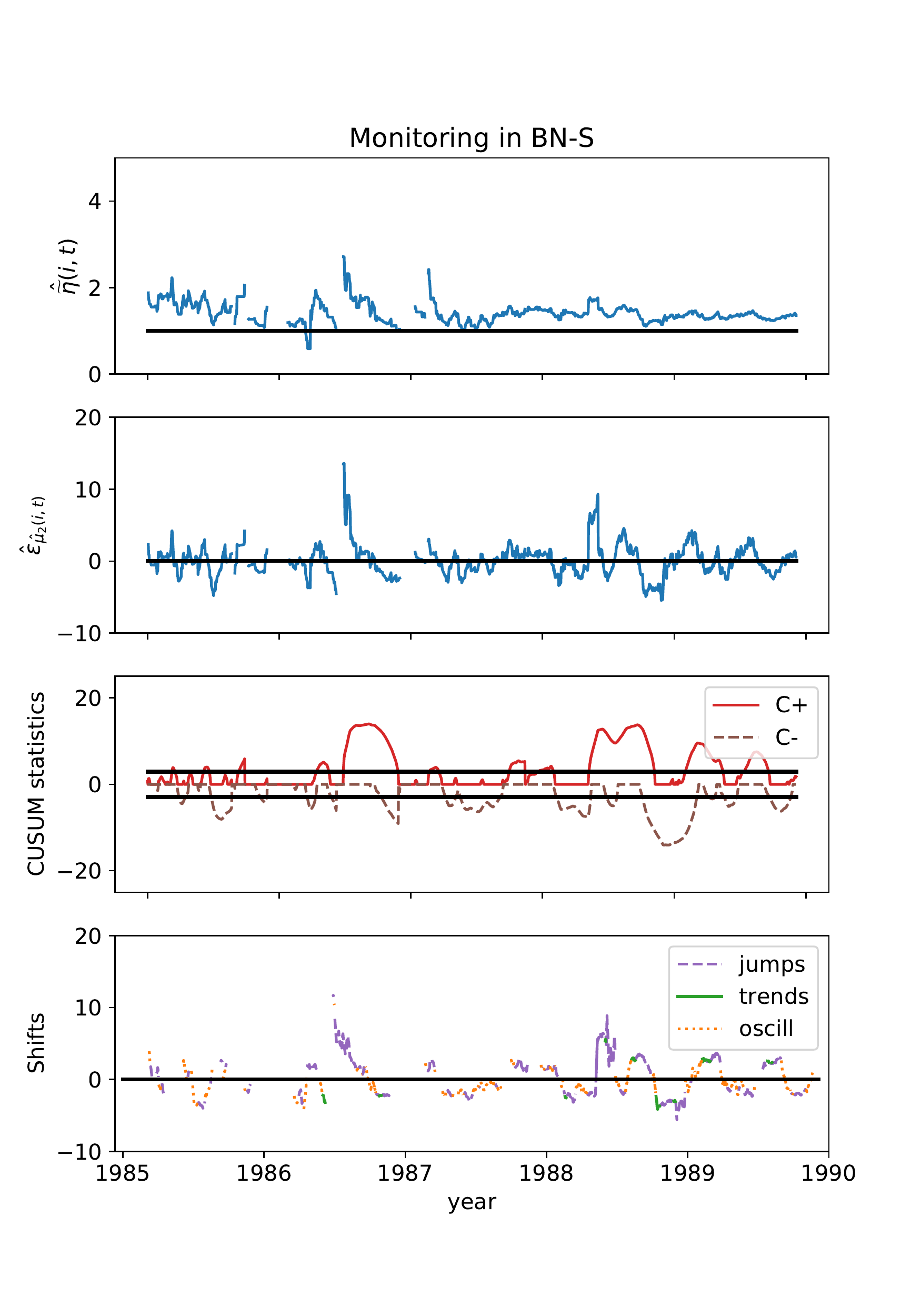}
		\caption{}
		\label{fig:BN-S}
	\end{subfigure}
\caption{\footnotesize{ (a) Control scheme applied on the data from the Athens observatory (A3) in Greece and (b) from the Wilhelm-Foerster-Sternwarte observatory (BN-S) in Berlin over 1985-1989. A3 has around 69\% of out-of-control observations and is included in the moderately stable pool, $P_1$, while BN-S is in alert around 75\% of its observing period (and is not included in $P_1$).}}
\end{figure}

\begin{figure}[!htb]
	\centering
	\begin{subfigure}{0.44\textwidth}
		\centering
		\includegraphics[scale=0.41]{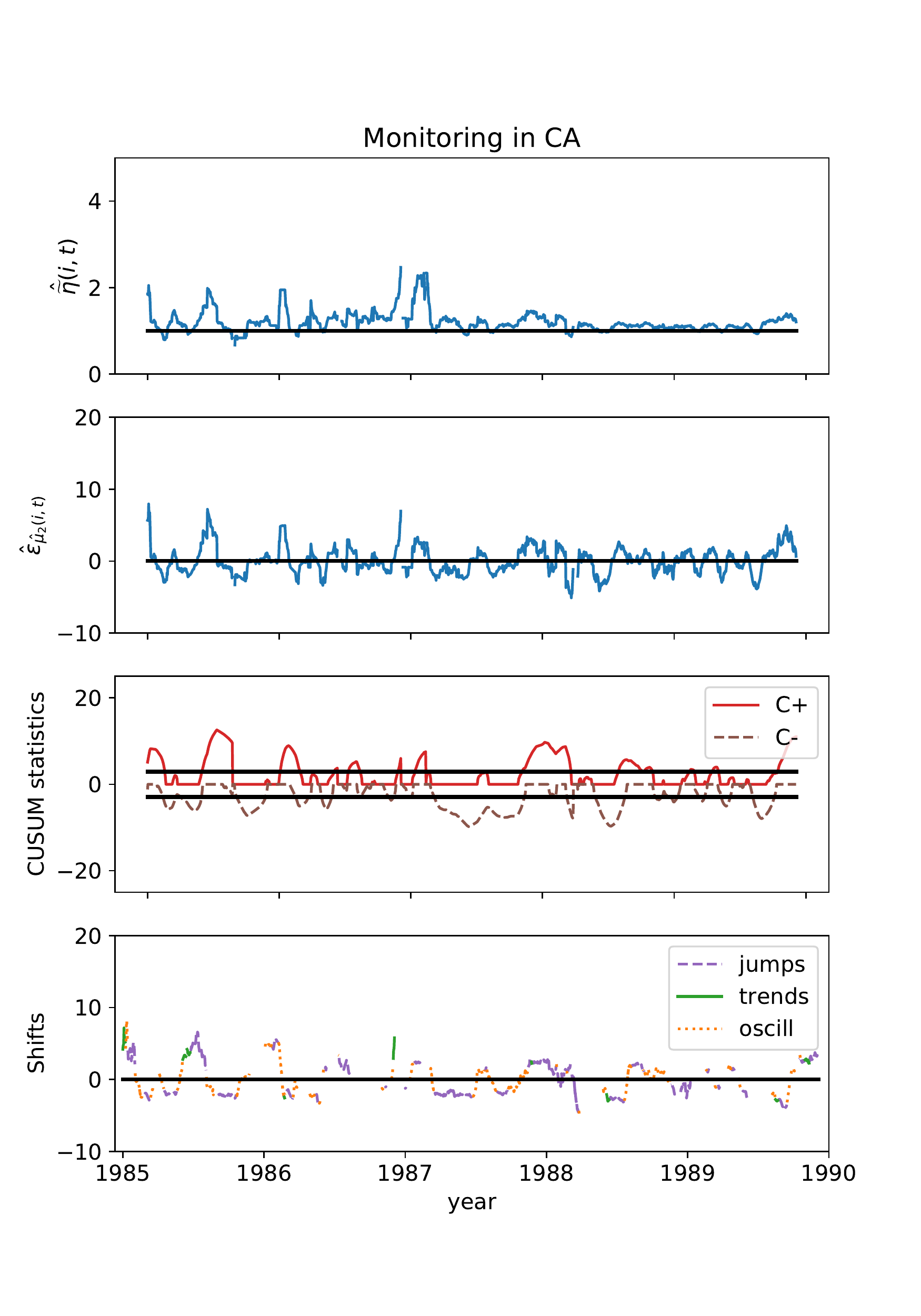}
		\caption{}
		\label{fig:CA}
	\end{subfigure}
	\begin{subfigure}{0.44\textwidth}
		\centering
		\includegraphics[scale=0.41]{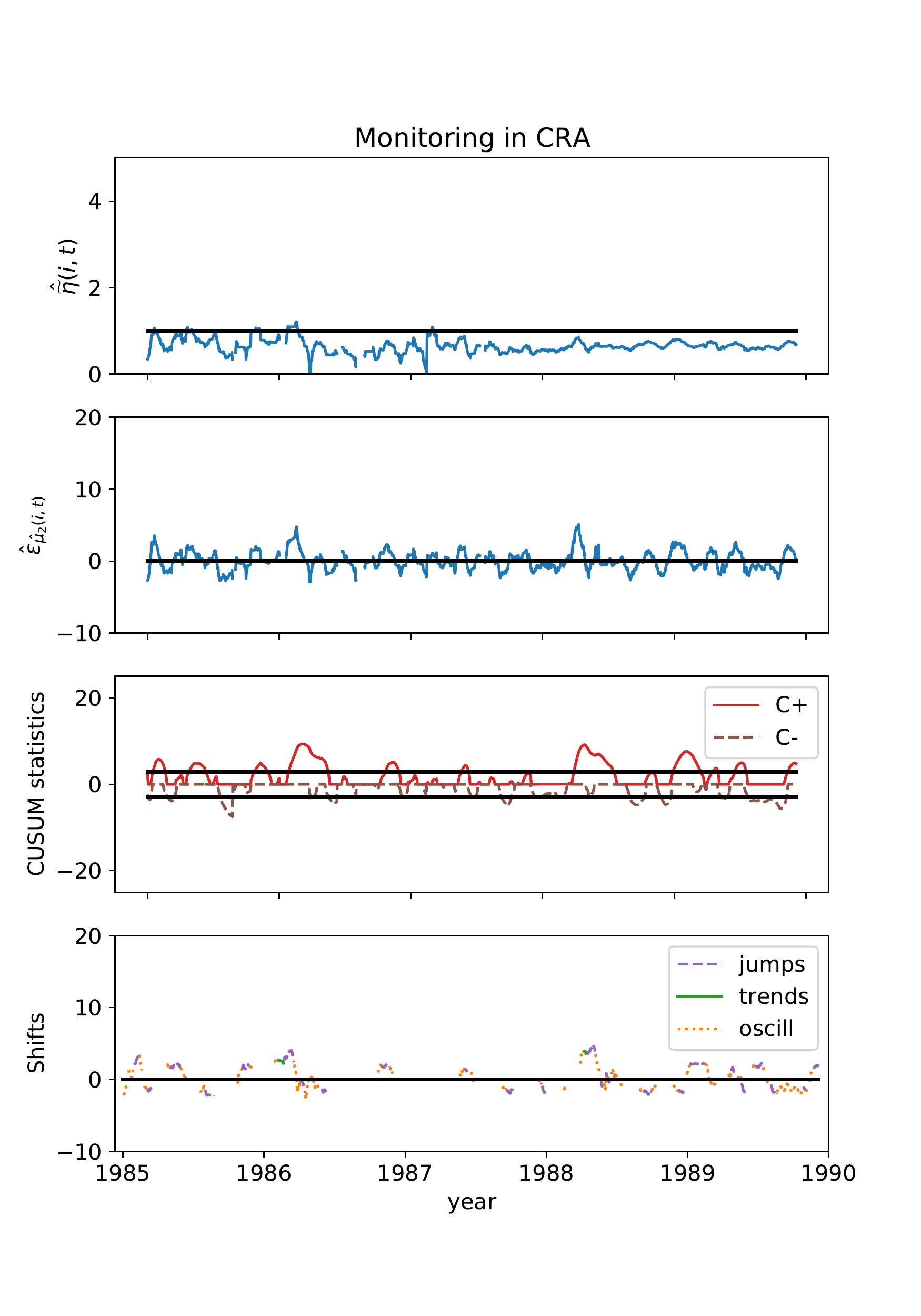}
		\caption{}
		\label{fig:CRA}
	\end{subfigure}
\caption{\footnotesize{ (a) Control scheme applied on the data from the Catania observatory (CA) in Italy and (b) from observer Thomas A. Cragg (CRA), who looked at the Sun in Australia over 1985-1989. CA has around 69\% of out-of-control observations and is included in $P_1$ while CRA is in alert around 53\% of its observing period and is included in the very stable pool, $P_2$.}}
\end{figure}

\begin{figure}[!htb]
	\centering
	\begin{subfigure}{0.44\textwidth}
		\centering
		\includegraphics[scale=0.41]{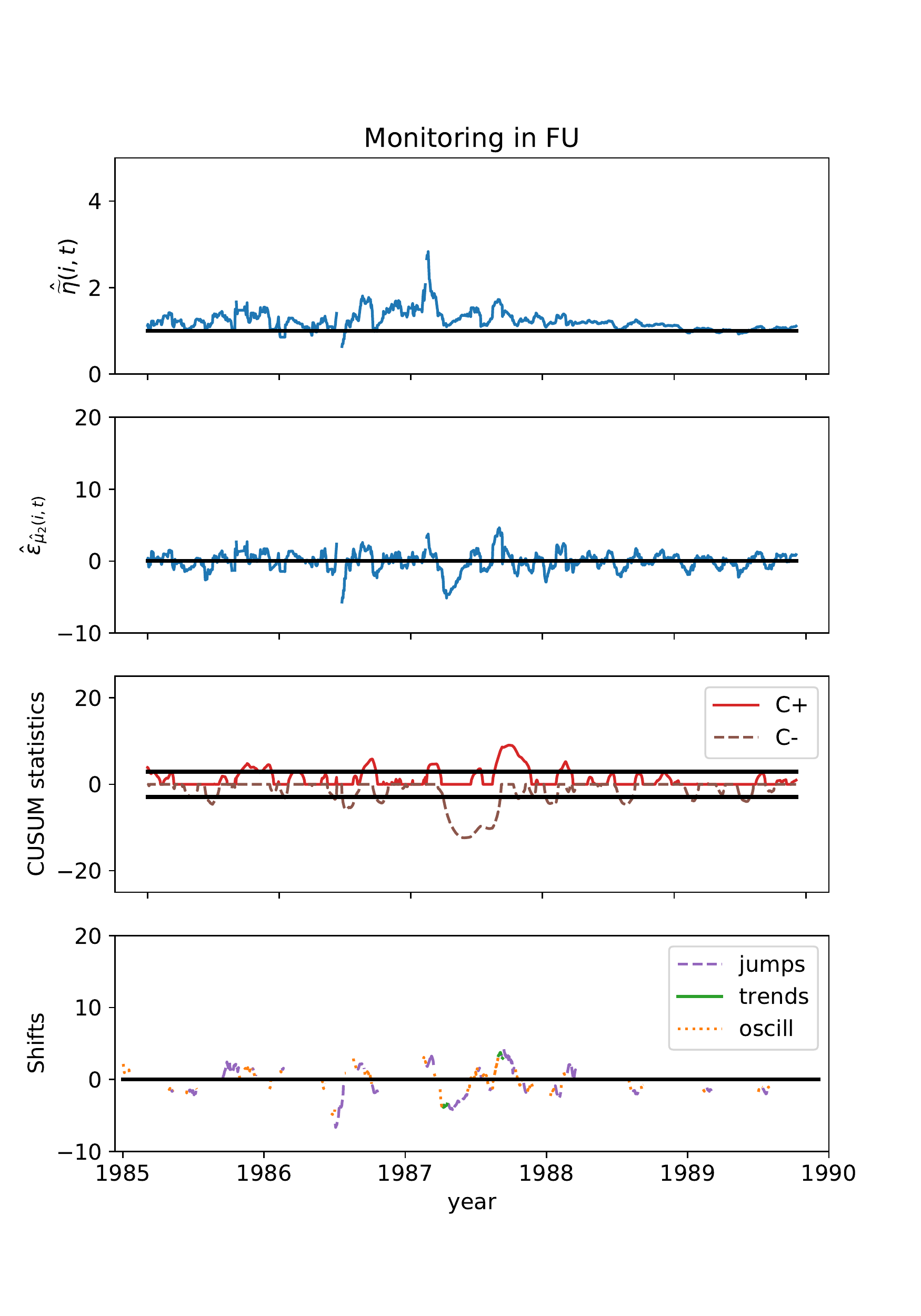}
		\caption{}
		\label{fig:FU}
	\end{subfigure}
	\begin{subfigure}{0.44\textwidth}
		\centering
		\includegraphics[scale=0.41]{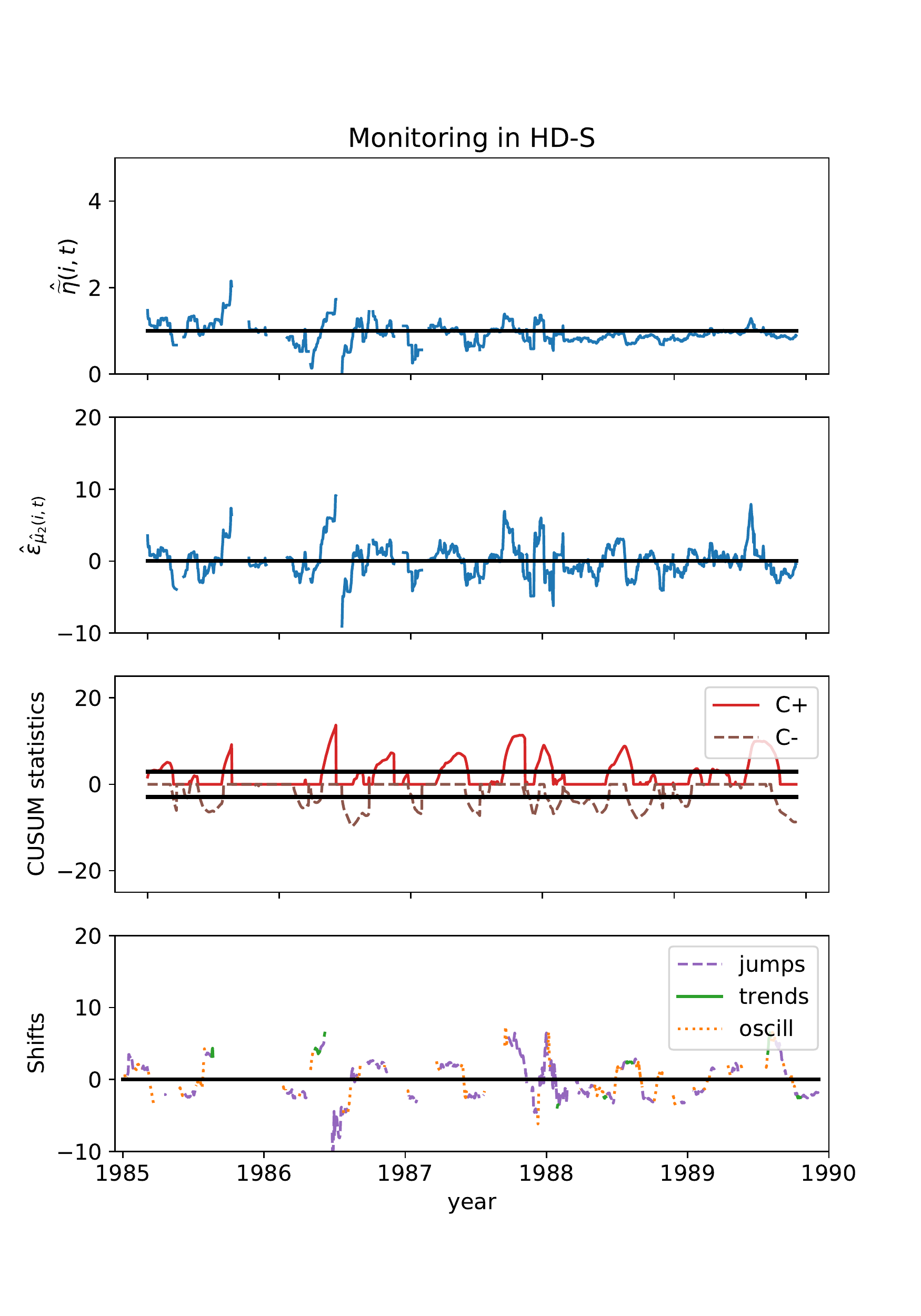}
		\caption{}
		\label{fig:HD-S}
	\end{subfigure}
\caption{\footnotesize{ (a)  Control scheme applied on the data from observer Fujimori-san (FU) located in Japan and (b) from observer Roland Hedewig (HD-S) based in Germany over 1985-1989. FU has around 41\% of out-of-control observations and is included in $P_2$ while HD-S is in alert around 72\% of its observing period. }}
\end{figure}

\begin{figure}[!htb]
	\centering
	\begin{subfigure}{0.44\textwidth}
		\centering
		\includegraphics[scale=0.41]{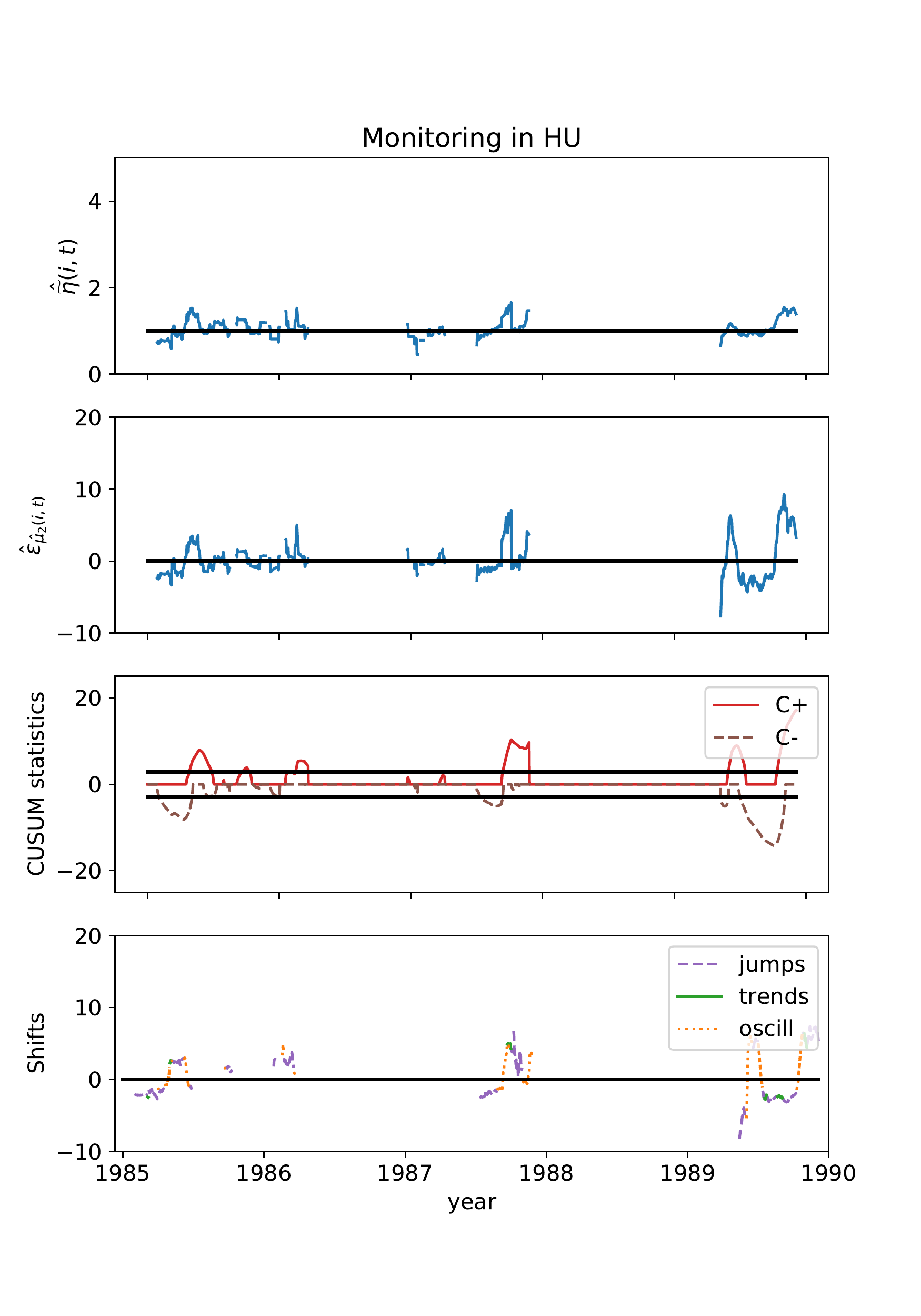}
		\caption{}
		\label{fig:HU}
	\end{subfigure}
	\begin{subfigure}{0.44\textwidth}
		\centering
		\includegraphics[scale=0.41]{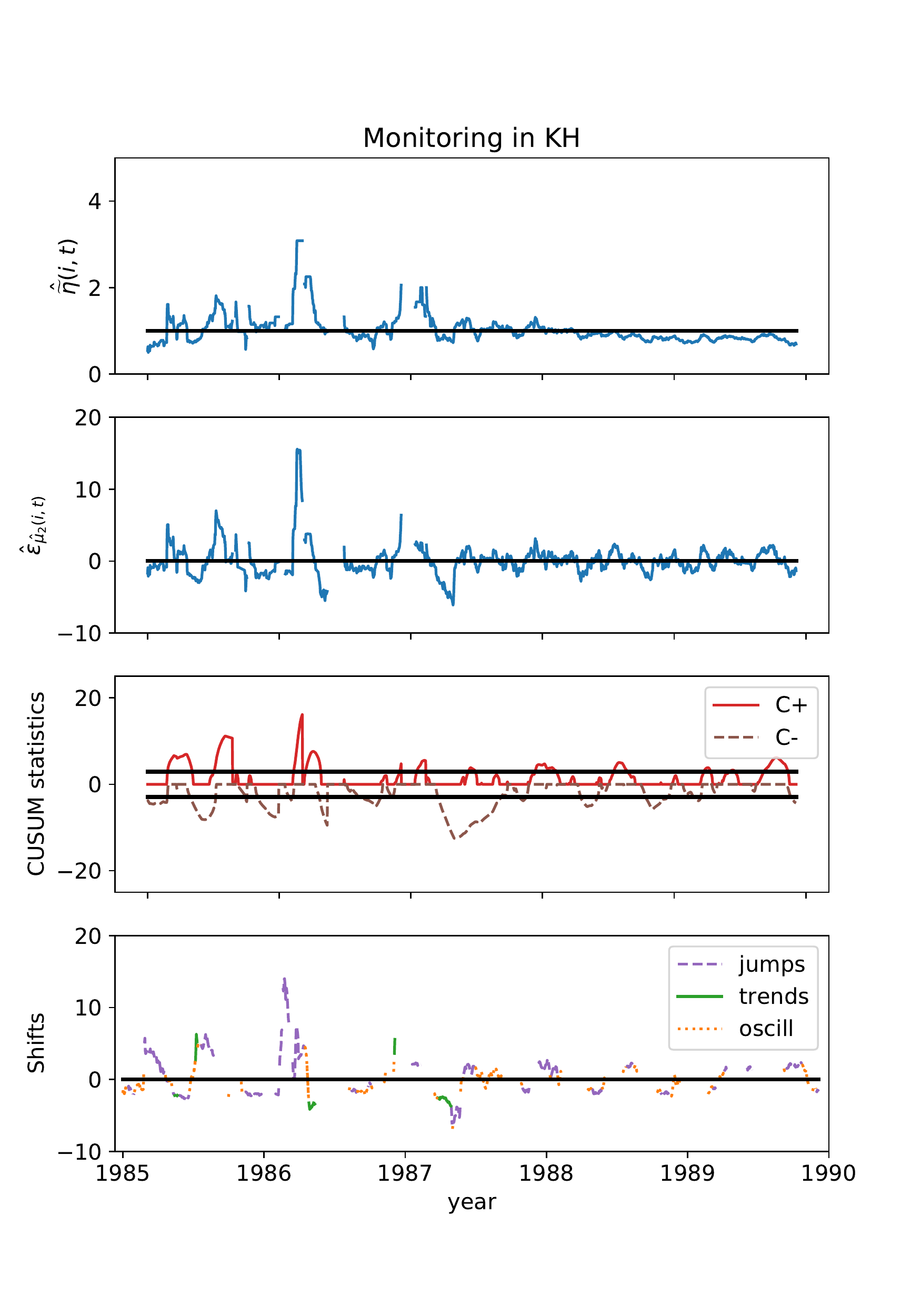}
		\caption{}
		\label{fig:KH}
	\end{subfigure}
\caption{\footnotesize{ (a) Control scheme applied on the data from the public observatory of Hurbanovo (HU) in Slovakia and (b) from the station Kandilli (KH) in Turkey over 1985-1989. HU has around 74\% of out-of-control observations while KH is in alert around 55\% of its observing period and is included in $P_2$. }}
\end{figure}

\begin{figure}[!htb]
	\centering
	\begin{subfigure}{0.44\textwidth}
		\centering
		\includegraphics[scale=0.41]{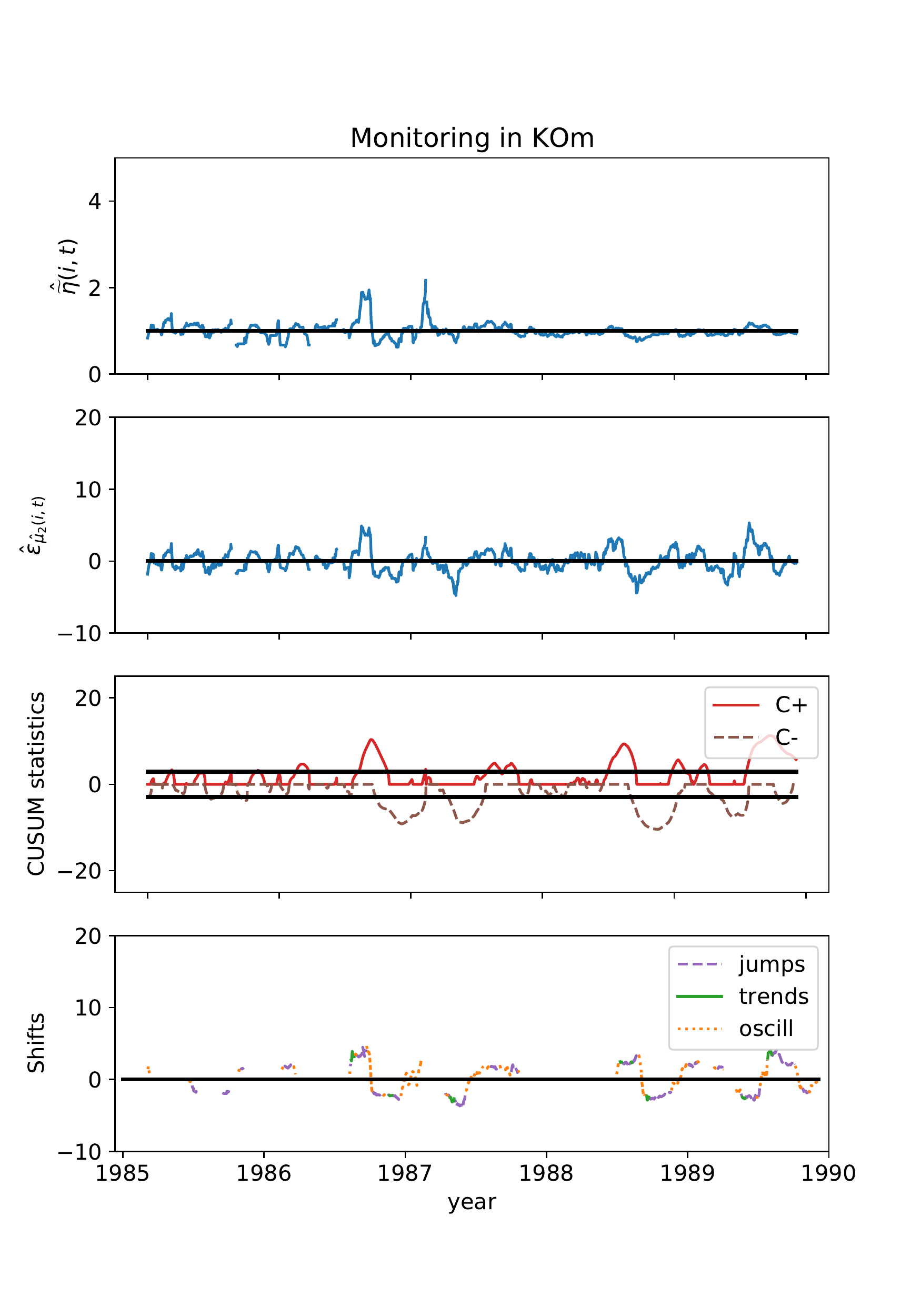}
		\caption{}
		\label{fig:KOm}
	\end{subfigure}
	\begin{subfigure}{0.44\textwidth}
		\centering
		\includegraphics[scale=0.41]{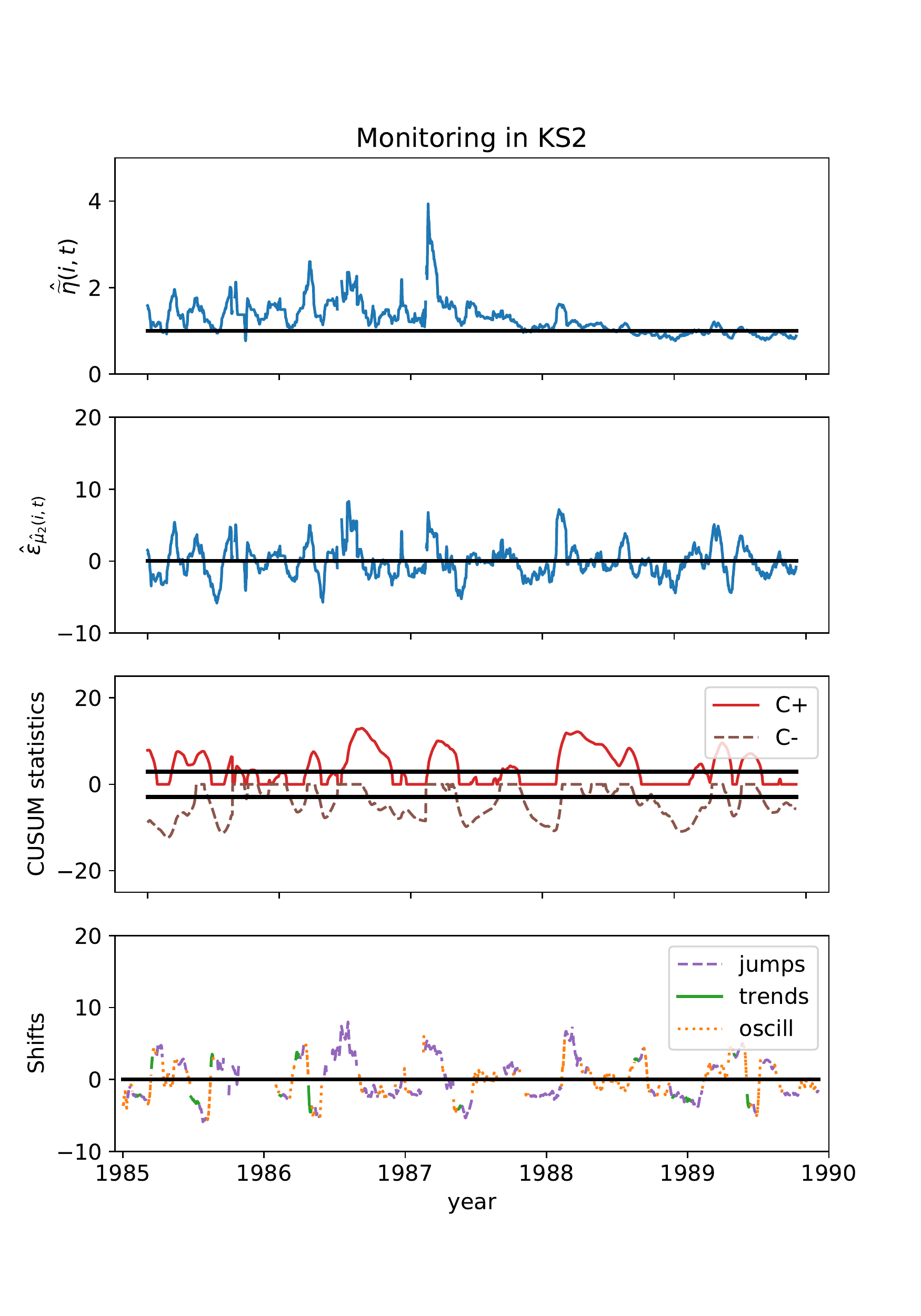}
		\caption{}
		\label{fig:KS2}
	\end{subfigure}
\caption{\footnotesize{ (a) Control scheme applied on the data from observer Koyama-san (KOm) located in Japan and (b) from the station Kislovodsk (KS2) in Russia over 1985-1989. KOm has around 69\% of out-of-control observations and is included in $P_1$ while KS2 is in alert around 83\%. }}
\end{figure}

\begin{figure}[!htb]
	\centering
	\begin{subfigure}{0.44\textwidth}
		\centering
		\includegraphics[scale=0.41]{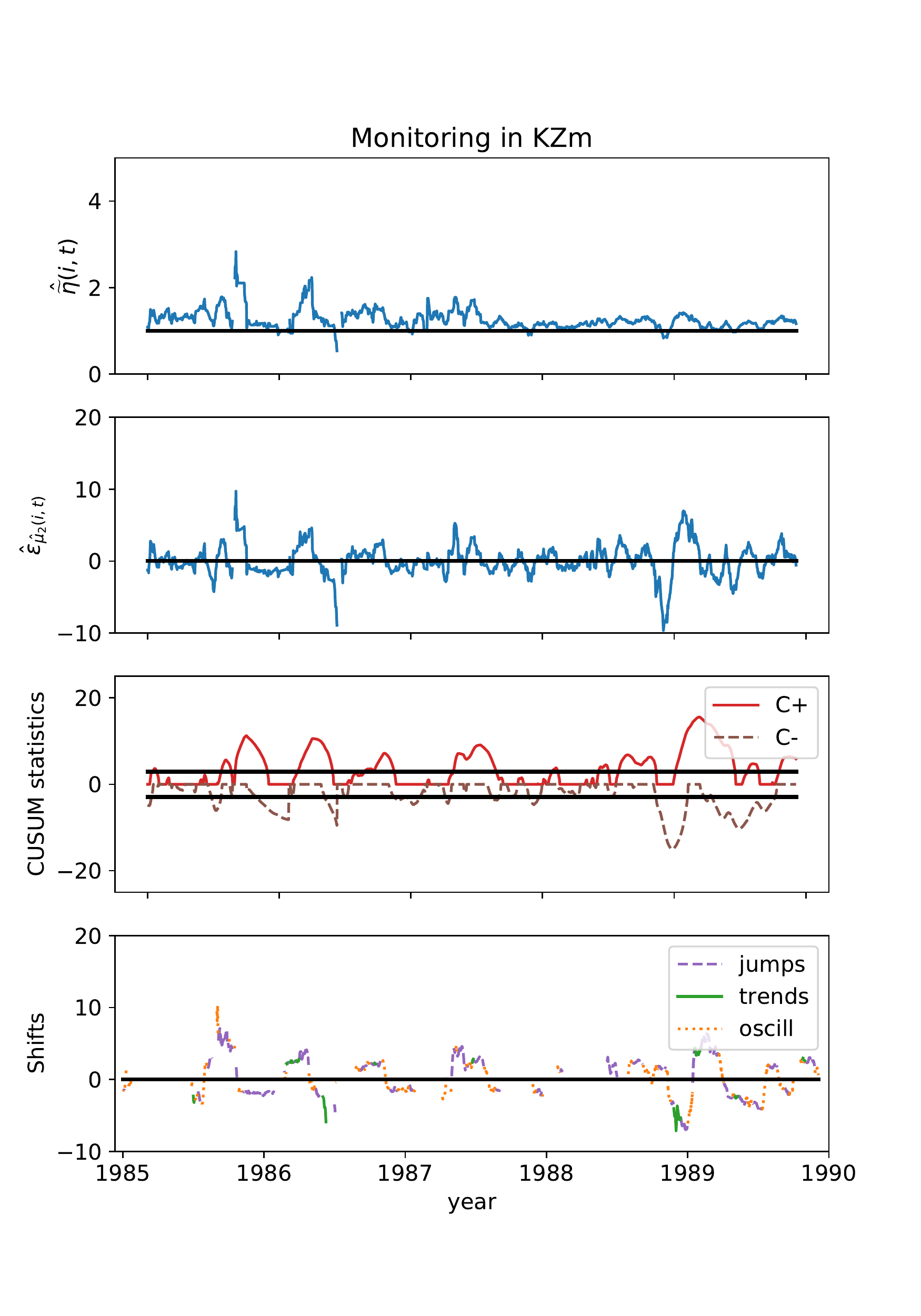}
		\caption{}
		\label{fig:KZm}
	\end{subfigure}
	\begin{subfigure}{0.44\textwidth}
		\centering
		\includegraphics[scale=0.41]{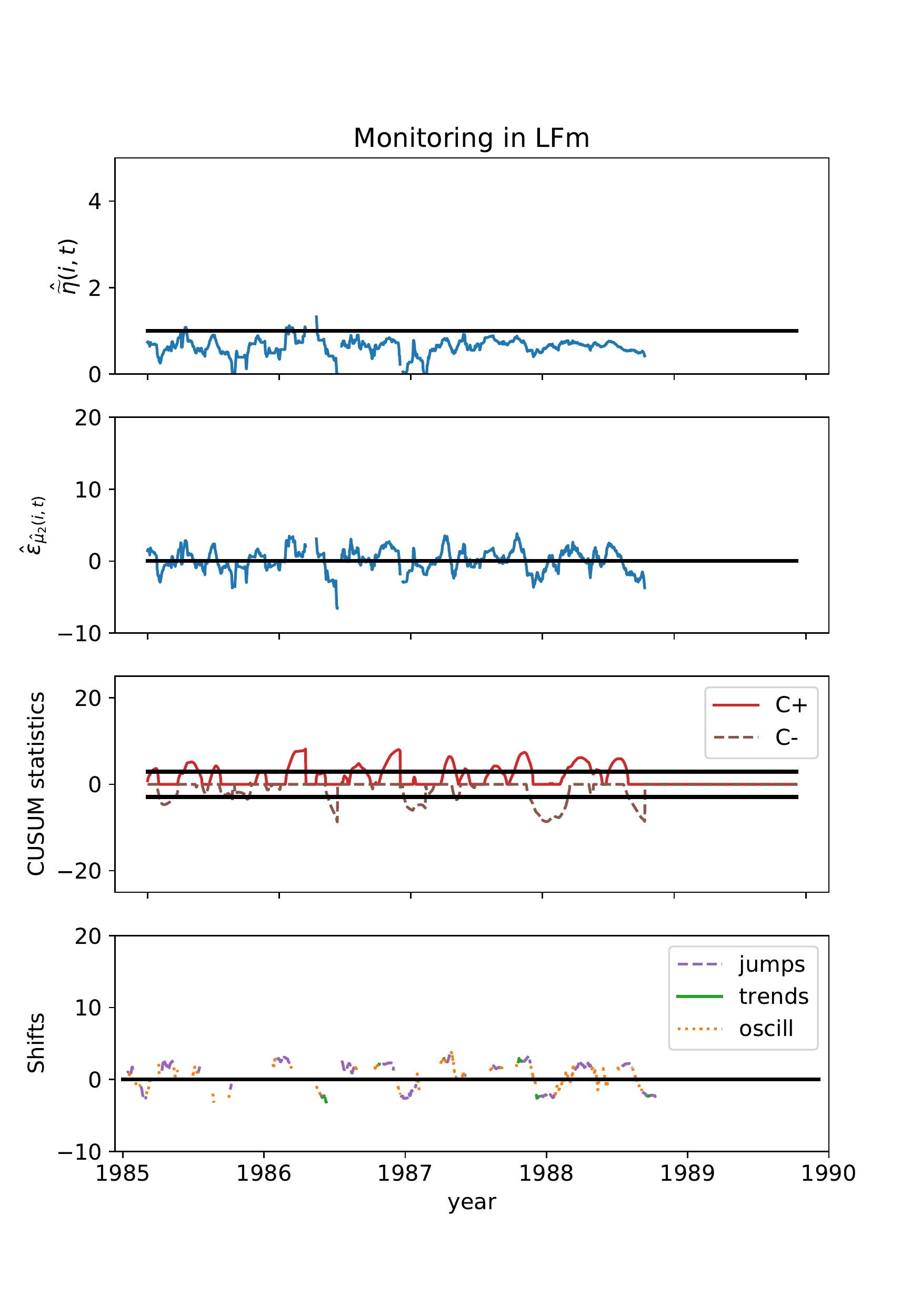}
		\caption{}
		\label{fig:LFm}
	\end{subfigure}
\caption{\footnotesize{ (a) Control scheme applied on the data from the station Kanzelhohe (KZm) in Austria and (b) from observer Herbert Luft (LFm) in New-York over 1985-1989. KZm has around 73\% of out-of-control observations while LFm is in alert around 71\% of its observing period and is included in $P_1$. }}
\end{figure}

\begin{figure}[!htb]
		\centering
		\includegraphics[scale=0.81]{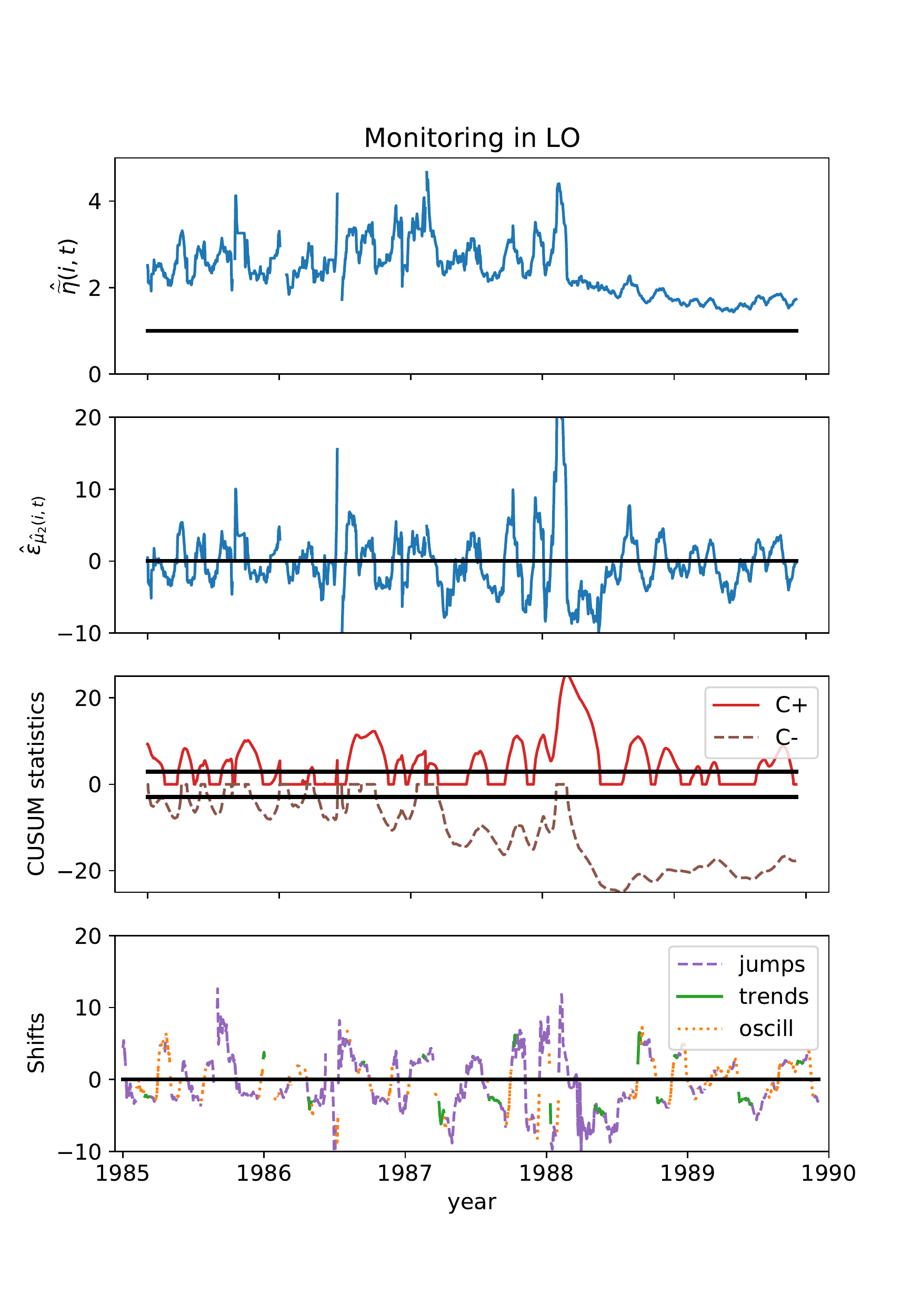}
		\caption{\footnotesize{ Control scheme applied on the data from the Specola Solare observatory of Locarno (LO) in Switzerland over 1985-1989. LO has around 90\% of out-of-control observations.}}
		\label{fig:LO}
\end{figure}

\begin{figure}[!htb]
	\centering
	\begin{subfigure}{0.44\textwidth}
		\centering
		\includegraphics[scale=0.41]{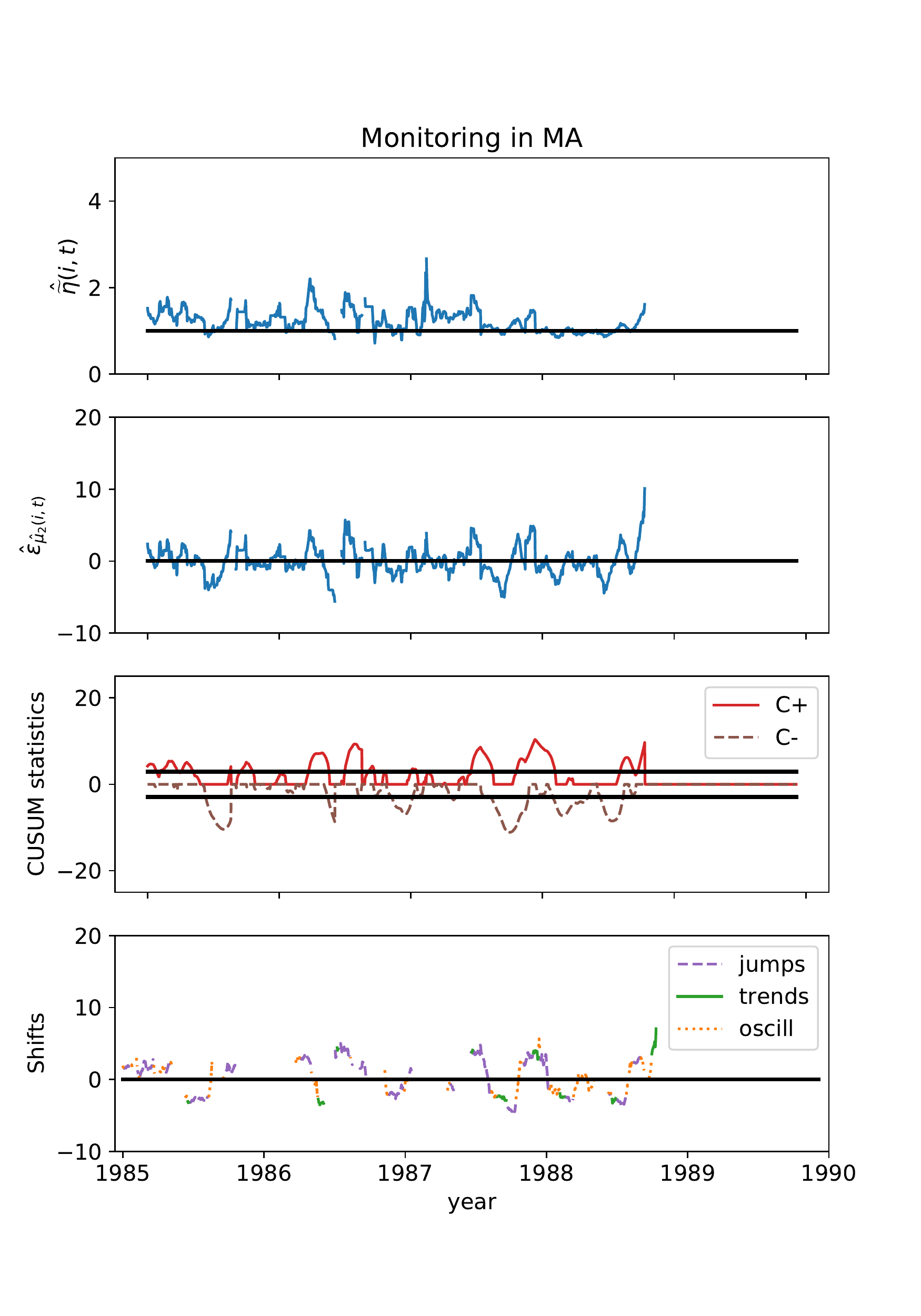}
		\caption{}
		\label{fig:MA}
	\end{subfigure}
	\begin{subfigure}{0.44\textwidth}
		\centering
		\includegraphics[scale=0.41]{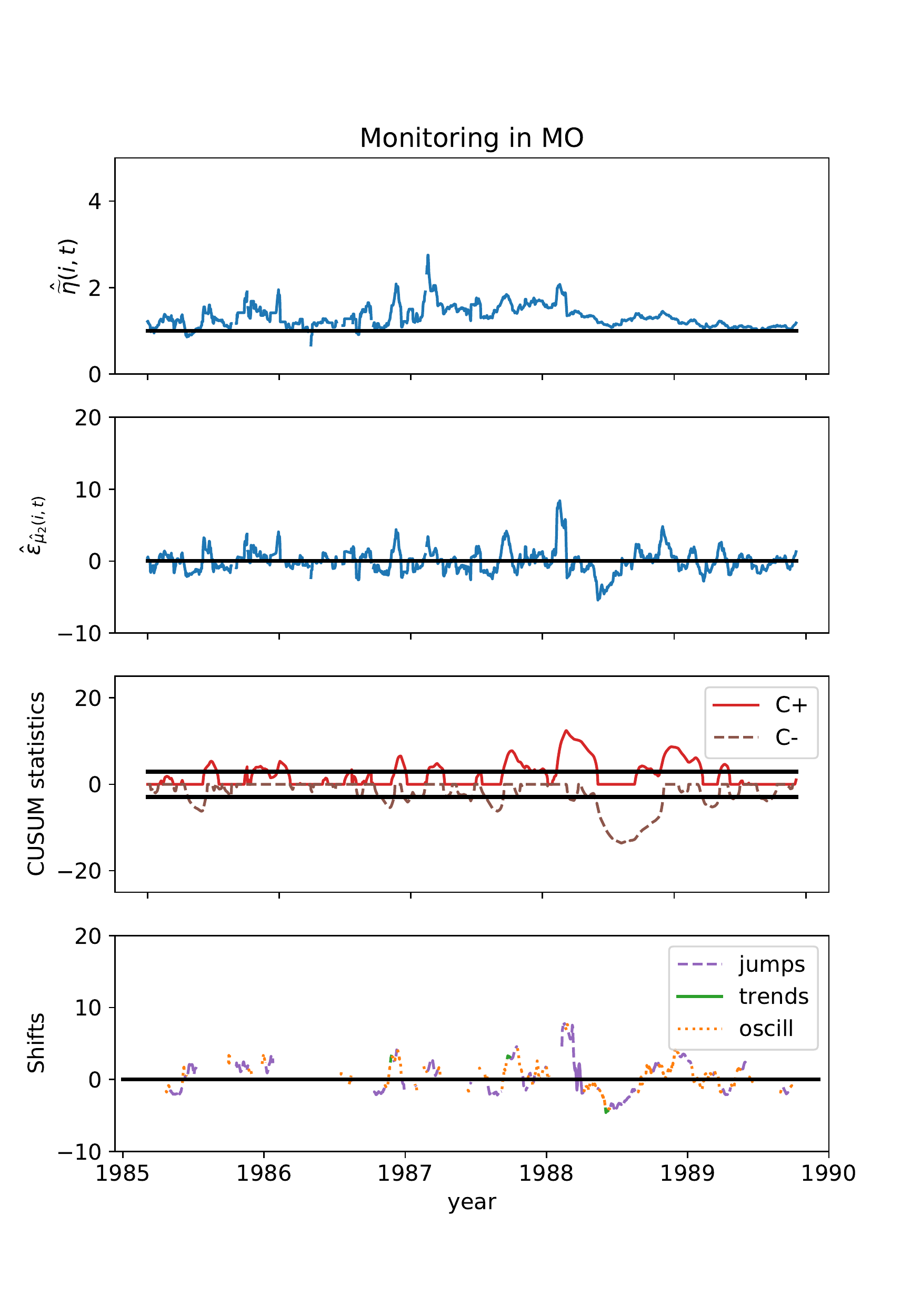}
		\caption{}
		\label{fig:MO}
	\end{subfigure}
\caption{\footnotesize{ (a) Control scheme applied on the data from the Manilla observatory (MA) in Philippines and (b) from observer Mochizuki-san (MO) located in Japan over 1985-1989. MA has around 66\% of out-of-control observations and is included in $P_1$ while MO is in alert around 59\% of its observing period and is included in $P_2$. }}
\end{figure}

\begin{figure}[!htb]
	\centering
	\begin{subfigure}{0.44\textwidth}
		\centering
		\includegraphics[scale=0.41]{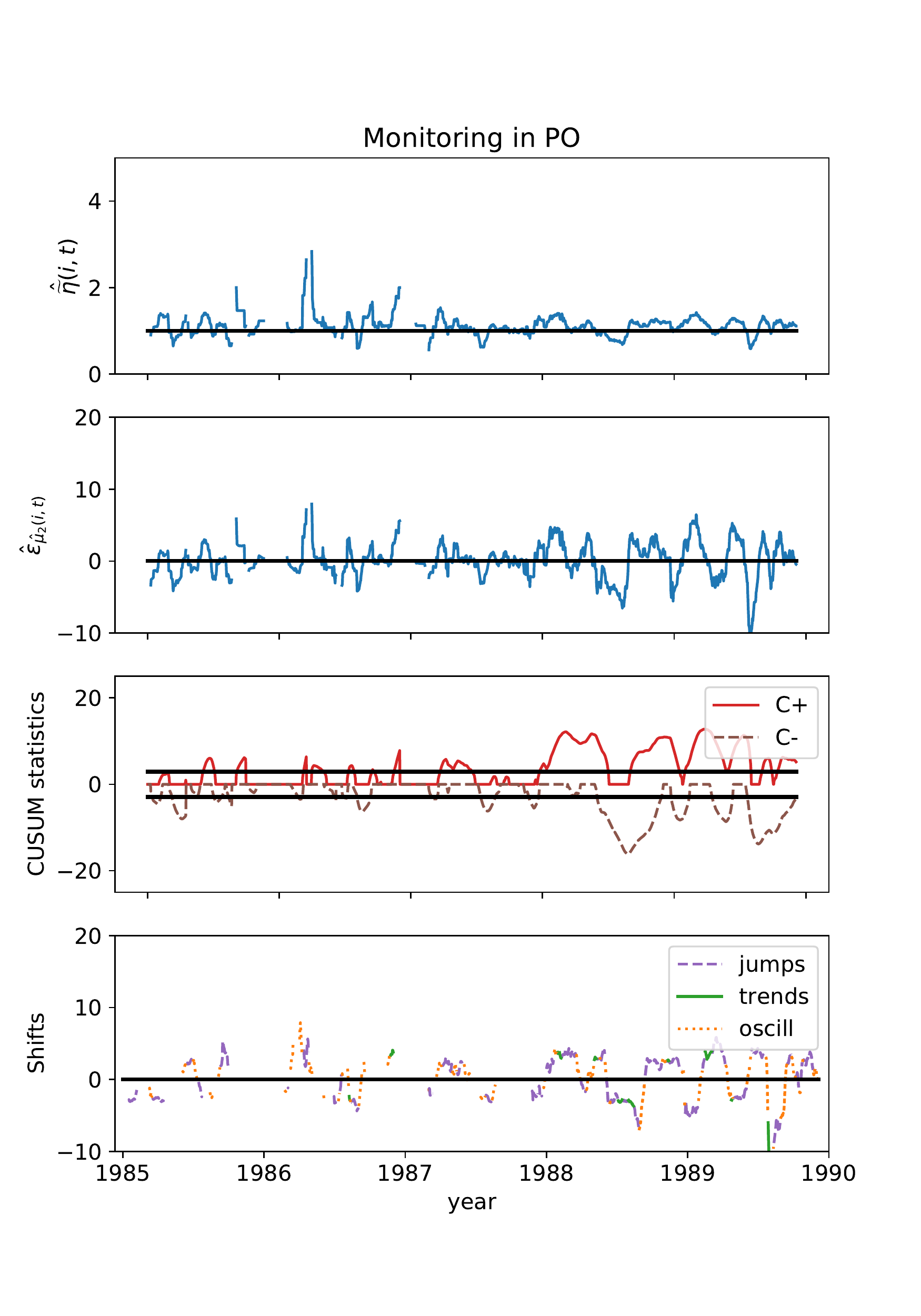}
		\caption{}
		\label{fig:PO}
	\end{subfigure}
	\begin{subfigure}{0.44\textwidth}
		\centering
		\includegraphics[scale=0.41]{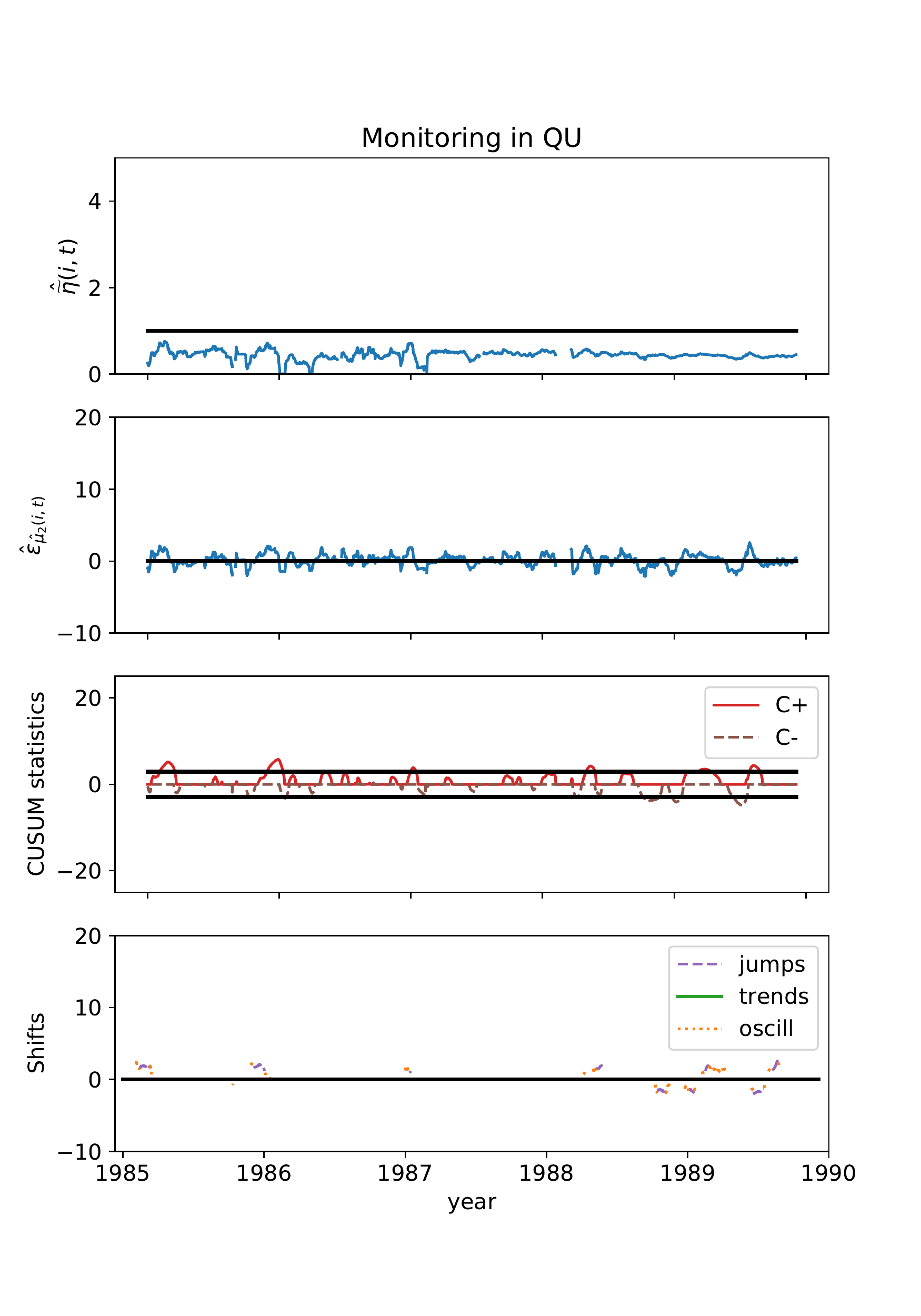}
		\caption{}
		\label{fig:QU}
	\end{subfigure}
\caption{\footnotesize{ (a) Control scheme applied on the data from the Postdam observatory (PO) in Berlin and (b) from the station Quezon (QU) in Philippines over 1985-1989. PO has around 68\% of out-of-control observations while QU is in alert around 40\% of its observing period and is included in $P_2$. }}
\end{figure}

\begin{figure}[!htb]
	\centering
	\begin{subfigure}{0.44\textwidth}
		\centering
		\includegraphics[scale=0.41]{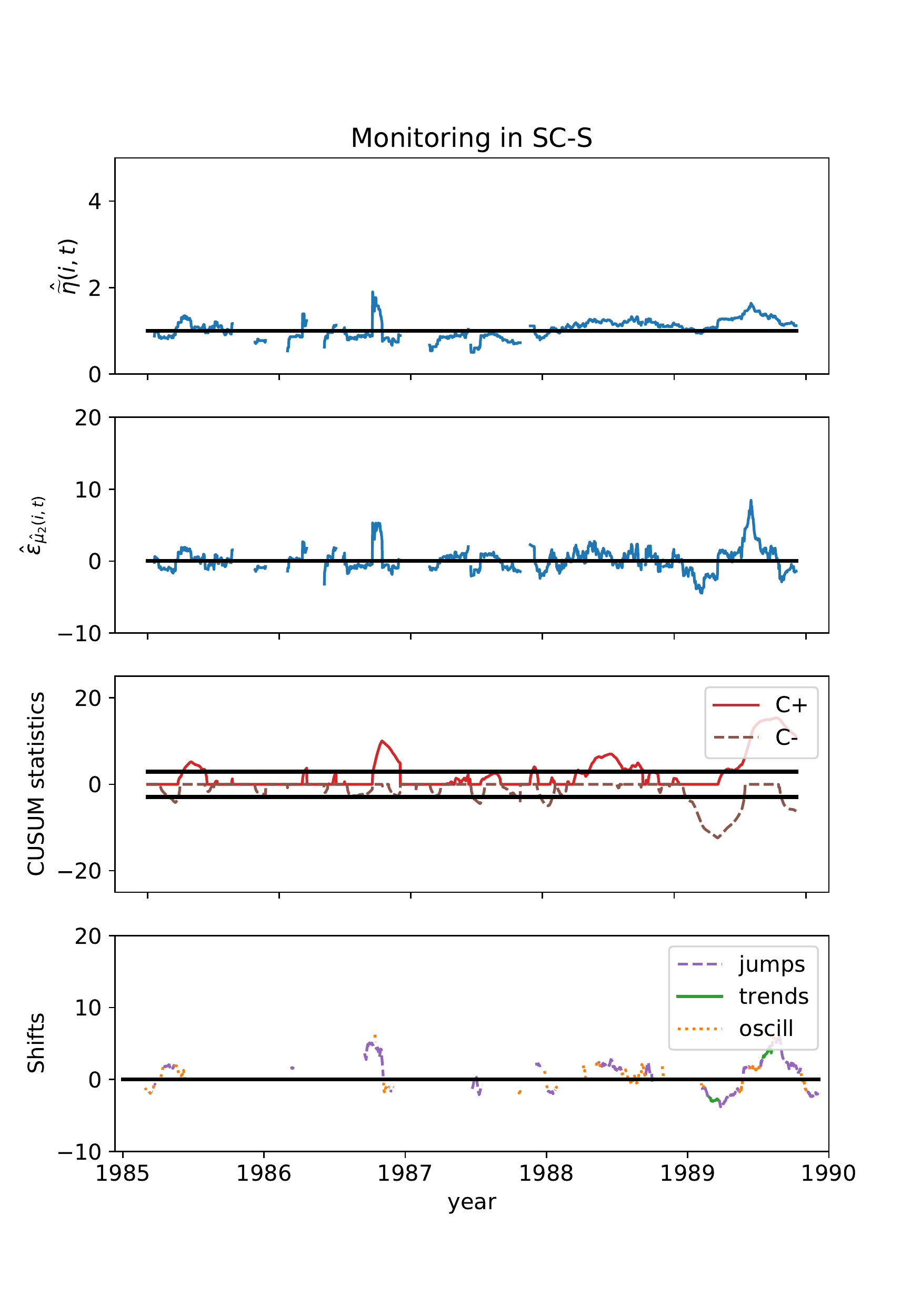}
		\caption{}
		\label{fig:SC-S}
	\end{subfigure}
	\begin{subfigure}{0.44\textwidth}
		\centering
		\includegraphics[scale=0.41]{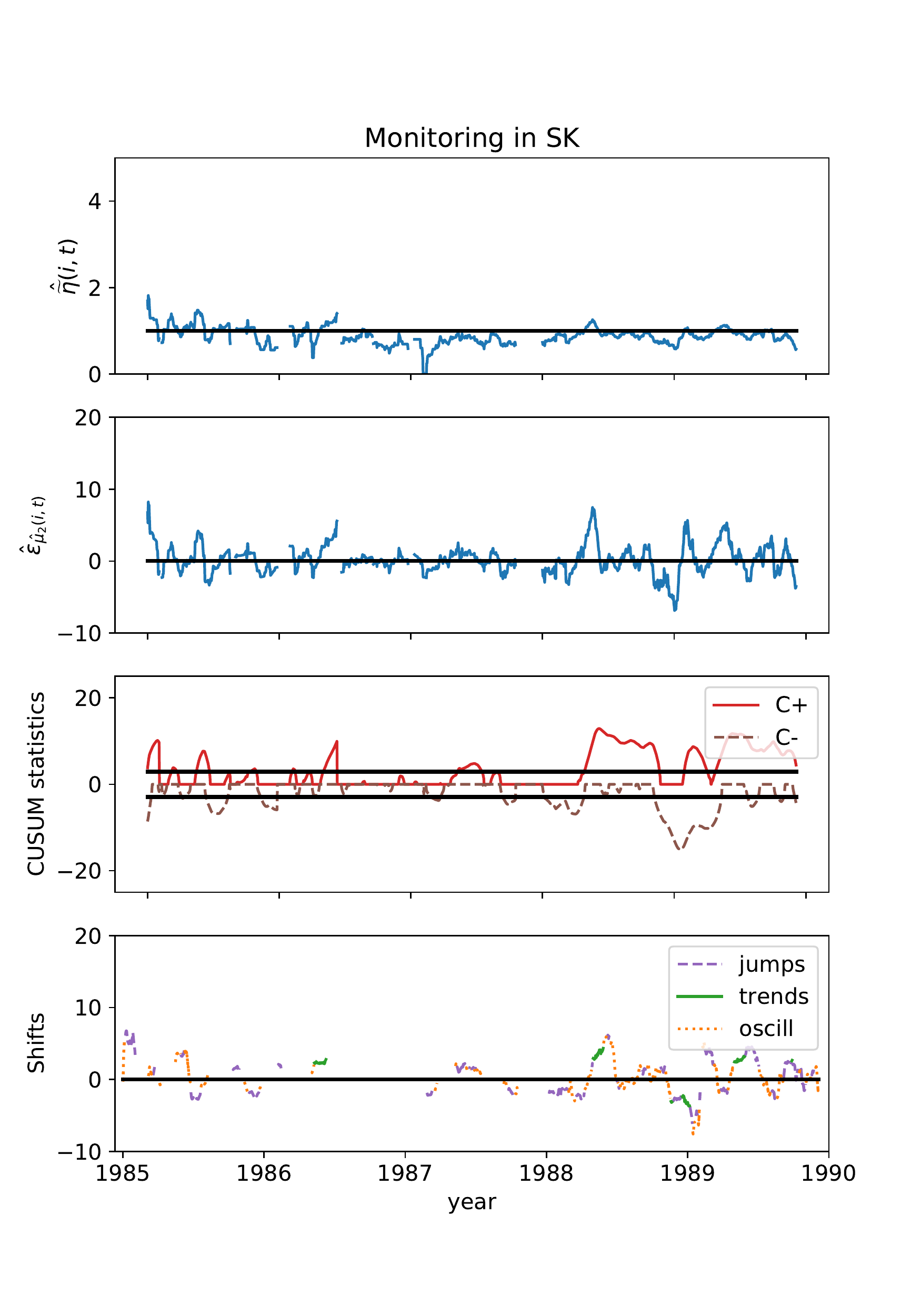}
		\caption{}
		\label{fig:SK}
	\end{subfigure}
\caption{\footnotesize{ (a) Control scheme applied on the data from observer Hubertus Schulze-Neuhoff (SC-S) located in Germany and (b) from the Skalnate Pleso observatory (SK) in Slovakia over 1985-1989. SC-S has around 54\% of out-of-control observations and is included $P_2$ while SK is in alert around 81\% of its observing period. }}
\end{figure}

\begin{figure}[!htb]
	\centering
	\begin{subfigure}{0.44\textwidth}
		\centering
		\includegraphics[scale=0.41]{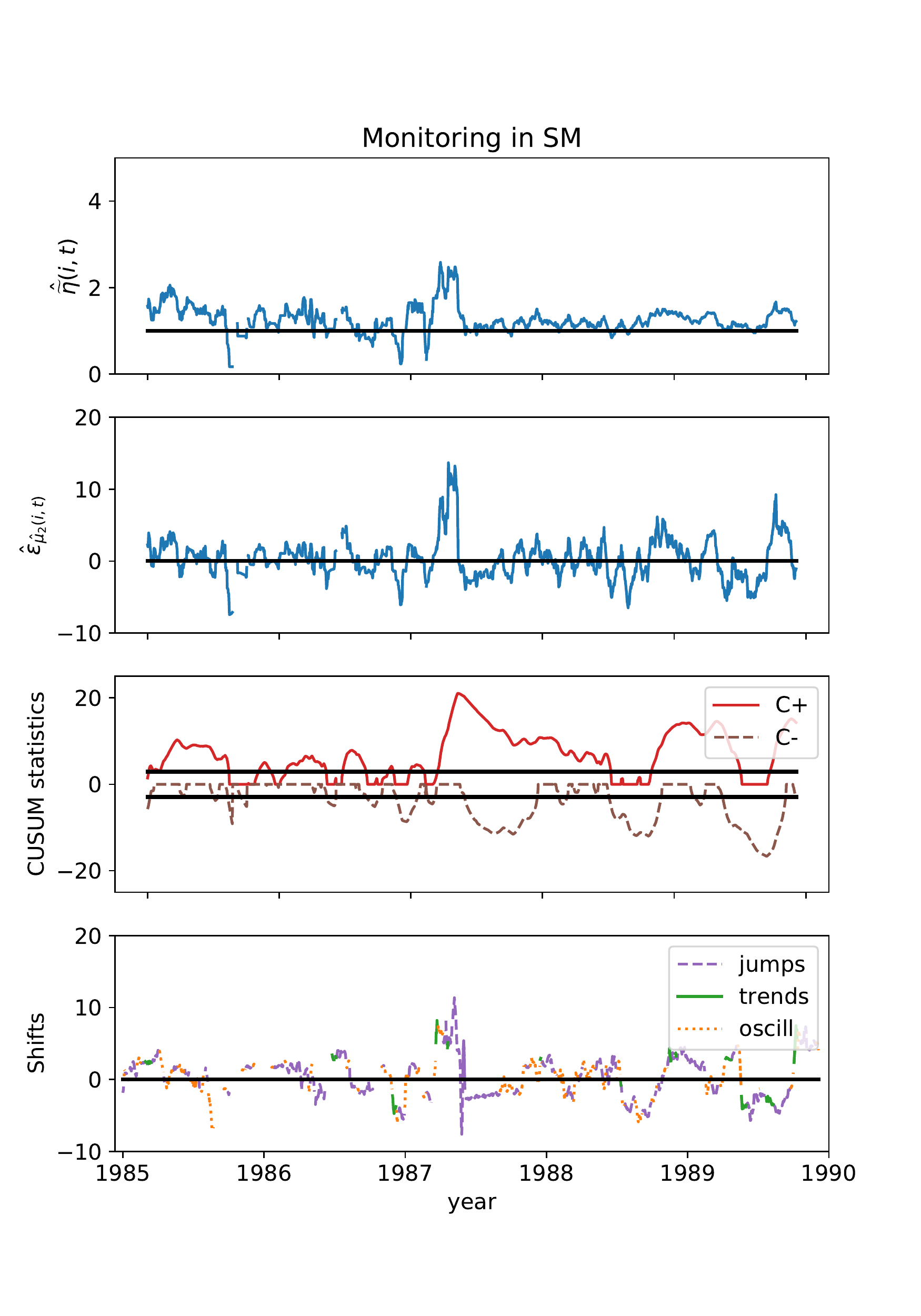}
		\caption{}
		\label{fig:SM}
	\end{subfigure}
	\begin{subfigure}{0.44\textwidth}
		\centering
		\includegraphics[scale=0.41]{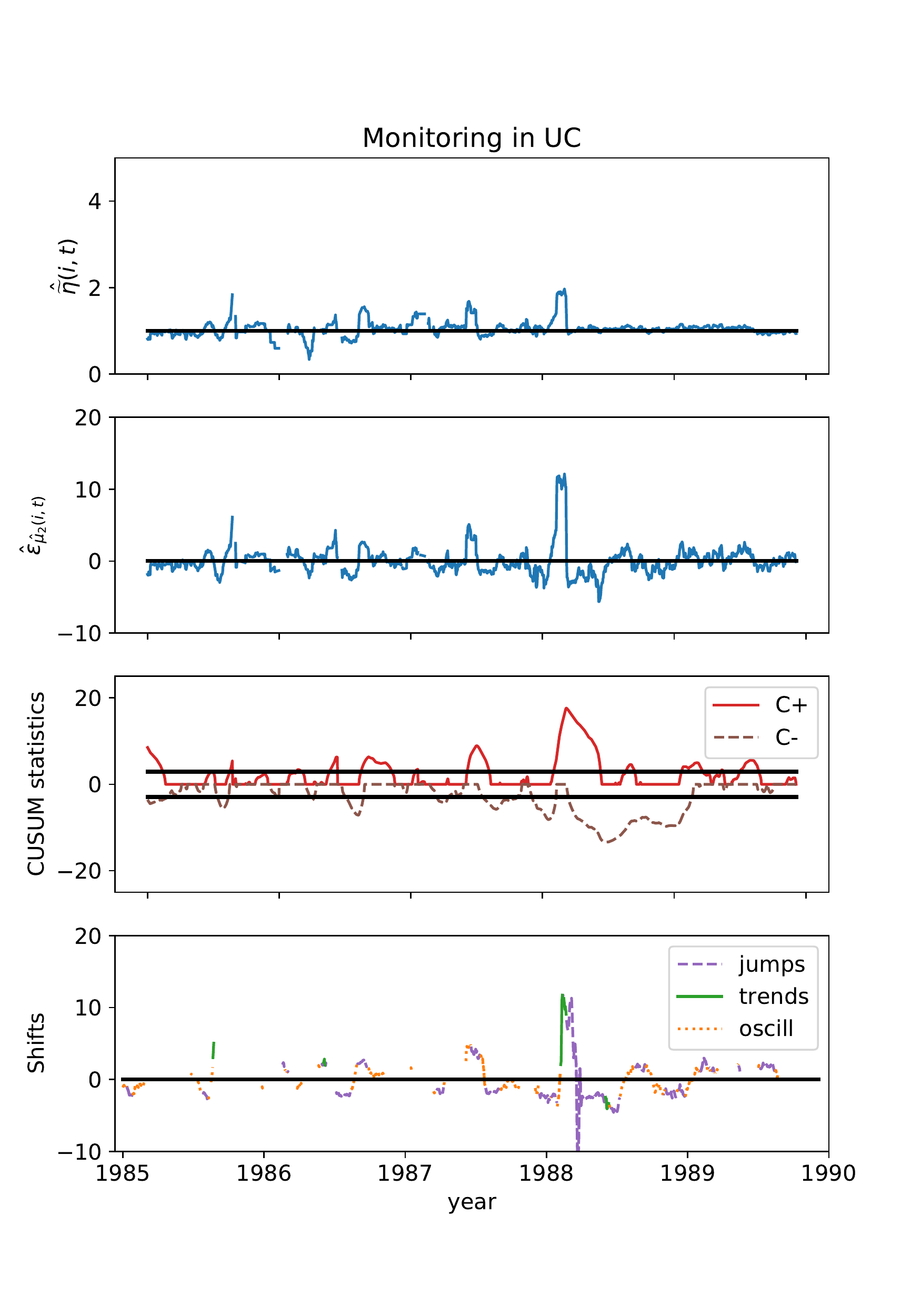}
		\caption{}
		\label{fig:UC}
	\end{subfigure}
\caption{\footnotesize{ (a) Control scheme applied on the data from the San Miguel observatory (SM) in Argentina and (b) from the USET station (UC) in Belgium over 1985-1989. SM has around 95\% of out-of-control observations while UC is in alert around 72\% of its observing period and is included in $P_1$. }}
\end{figure}

\end{document}